\documentclass[twocolumn]{aastex631}

\usepackage{graphicx}	
\usepackage{amsmath,bm,amssymb}	
\usepackage{multirow, threeparttable, booktabs, makecell}
\usepackage{hyperref}

\newcommand{\njk}{N_{\rm JK}}

\newcommand{\ZY}[1]{{\color{black} #1}}
\newcommand{\LSH}[1]{{\color{black} #1}}

\begin{document}

\title[Joint Cosmology with GC & kSZ]{\boldmath First Cosmological Constraints from the Joint Analysis of Galaxy Clustering and the Kinetic Sunyaev-Zel’dovich Effect}
\correspondingauthor{Yi Zheng}
\email{zhengyi27@mail.sysu.edu.cn}

\author{Shaohong Li}
\affiliation{School of Physics and Astronomy, Sun Yat-sen University, 2 Daxue Road, Tangjia, Zhuhai, 519082, People’s Republic of China}
\affiliation{CSST Science Center for the Guangdong–Hong Kong–Macau Greater Bay Area, SYSU, Zhuhai, 519082, People’s Republic of China}
\author{Yi Zheng}%
\affiliation{School of Physics and Astronomy, Sun Yat-sen University, 2 Daxue Road, Tangjia, Zhuhai, 519082, People’s Republic of China}
\affiliation{CSST Science Center for the Guangdong–Hong Kong–Macau Greater Bay Area, SYSU, Zhuhai, 519082, People’s Republic of China}

\begin{abstract}
We perform the first joint analysis of galaxy clustering (GC) and the kinetic Sunyaev–Zel’dovich (kSZ) effect to simultaneously constrain cosmological and astrophysical parameters in this work, utilizing a combination of the Atacama Cosmology Telescope (ACT) Data Release 6 (DR6) map and the Constant Stellar Mass (CMASS) galaxy sample. As a complementary probe to the galaxy density power spectrum, we incorporate the pairwise kSZ power spectrum detected with a high signal-to-noise ratio (S/N $\sim 7$) to derive constraints on cosmological parameters ({$H_0 = 70.82^{+4.94}_{-5.01}$, $\Omega_{\rm m} = 0.290^{+0.092}_{-0.068}$, $w_0 = -1.038^{+0.245}_{-0.437}$}) and the average optical depth of the galaxy sample ($\lg\bar{\tau} = -4.24 \pm 0.10$). Compared to the GC-only analysis, the joint analysis yields tighter constraints on these cosmological parameters: the Figures of Merit improve by 20.5\%, 19.7\% and 10.0\% for the $H_0$--$\Omega_{\rm m}$, $H_0$--$w_0$ and $\Omega_{\rm m}$--$w_0$ contours, respectively. For the first time, we demonstrate the complementary applicability of the kSZ effect in constraining cosmological parameters from real observational data.
\end{abstract}


\section{Introduction}

To explain the accelerated expansion of the Universe~\citep{Riess1998,Perlmutter1999}, numerous theoretical models incorporating  dark energy or modified gravity have been developed~\citep{Clifton2012,Brax2018}, which exhibit mutual degeneracies. Breaking this degeneracy requires simultaneously measuring the cosmic expansion and structure growth histories~\citep{Weinberg2013,Joyce2016,Koyama2016}. The expansion history can be measured by adopting distance measurement methods such as standard candles~\citep{Riess2022}, standard rulers~\citep{Eisenstein2005}, standard sirens~\citep{Abbott2017}, standard shapes~\citep{Alcock1979,LiXD2016}, time-delay techniques~\citep{Wong2020,Treu2022}, among others. Information about structure growth can be derived from weak-lensing phenomena~\citep{Bartelmann2001,Hoekstra2008,Kilbinger2015} and the cosmic peculiar velocity field~\citep{Hamilton1998}. 

Both histories can be probed through galaxy clustering (GC) analysis. The expansion history is measured via the Alcock-Paczyński (AP) effect~\citep{Alcock1979}, while the growth history is detected through redshift-space distortions (RSD)~\citep{Kaiser1987}. The RSD effects manifest as anisotropic GC in redshift space, induced by the cosmic peculiar velocity field~\citep{Peacock2001,Guzzo2008,Samushia2012,Alam2017,Gil-Marin2020}. The same velocity field also generates the kinetic Sunyaev--Zel'dovich (kSZ) effect, a secondary cosmic microwave background (CMB) anisotropy resulting from the inverse-Compton scattering of CMB photons off free electrons with bulk peculiar motion~\citep{kSZ1970,kSZ1972,kSZ1980}. In this work, we study the cosmological constraints derived from the synergy of these two complementary probes.

Multitracer joint analyses are key to overcoming cosmic variance, parameter degeneracies, and systematics in cosmology~\citep{Seljak2009,McDonald2009b,CaiYC2012}. The combination of GC and the kSZ effect exemplifies this synergy, providing independent and complementary constraints on the growth of structure that are crucial for next-generation surveys targeting dark energy and modified gravity~\citep{Sugiyama2017,ZhengY2020,Okumura2022,Xiao2023} for ongoing and future projects such as DESI~\citep{DESI2016}, PFS~\citep{Takada2014}, Euclid~\citep{Euclid2020}, and CSST~\citep{CSST2025}.

Despite this potential, current kSZ applications remain largely confined to studies of halo gas profiles and baryonic feedback, a focus dictated by the limited signal-to-noise ratio (S/N; $\sim 4$--10) of current detections~\citep[e.g.,][]{Soergel16,Schaan2021,Calafut2021,Kusiak2021,Chen2022,Schiappucci2023,Hadzhiyska2024,Li2024,Ried-Guachalla2025}. The advent of experiments like Simons Observatory (SO)~\citep{Simons2019} and CMB-S4~\citep{CMBS42019-2}, projecting S/N $\sim \mathcal{O}(100)$~\citep[e.g.,][]{Sugiyama2018,Smith2018,ZhengY2024}, will transform the kSZ effect into a powerful cosmological tool. In this work, we pioneer its use in a joint analysis with GC, thereby extending its application from astrophysical studies of baryons to rigorous cosmological tests. We expect this approach will ultimately evolve into a unified framework capable of simultaneously constraining cosmology and baryonic physics.

This Letter is organized as follows. Section~\ref{sec:data} describes the datasets used in this analysis. {Section~\ref{sec:methodology} provides a brief summary of the power spectrum measurement procedure.} Section~\ref{sec:theory} outlines the theoretical framework based on nonlinear perturbation theory. Section~\ref{sec:results} presents the resulting cosmological and astrophysical constraints. We conclude with a summary of our findings in Section~\ref{sec:conclusion}. Additional technical details, including the validation tests of the theoretical models using mock observations, are provided in the appendices.

\section{Data}
\label{sec:data}

\subsection{Atacama Cosmology Telescope map}
\label{subsec:ACT_map}
\begin{figure}[t!]
\includegraphics[width=0.45\textwidth]{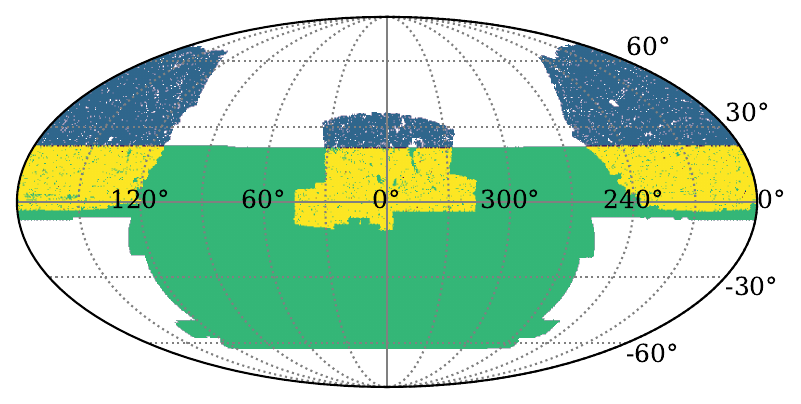}
\caption{\label{fig:sky_distribution} The sky coverage of the ACT map and CMASS galaxies with $N_{\rm side}=256$ in the HEALPix grid frame~\citep{Gorski2005}. The yellow areas represent the overlapping regions between the two data sets selected and used in this work. The purple ones are galaxies that have been removed due to masking. The blue and green pixels represent the remaining CMASS and ACT data, respectively.}
\end{figure}

The kSZ temperature signal is extracted from the arcminute-resolution CMB temperature map provided by the Atacama Cosmology Telescope (ACT) Data Release 6 (DR6)~\citep{Naess2025}\footnote{\href{https://lambda.gsfc.nasa.gov/product/act/act_dr6.02/act_dr6.02_maps_coadd_get.html}{act-planck\_dr4dr6\_coadd\_AA\_daynight\_f150\_map\_srcfree.fits}}. We utilize the combined day-night map at 150 GHz (f150) with point sources removed, which is a coaddition of ACT DR4 and Planck data. This f150 map has an effective full width at half maximum (FWHM) of $1.42$ arcmin and a median noise level of $14\ \mu\text{K}\cdot\text{arcmin}$. The map is stored in a Plate Carr\'ee projection in equatorial coordinates, with a pixel grid of $43200 \times 10320$ pixels (each $0.5 \ \text{arcmin} \times 0.5 \ \text{arcmin}$), covering the region $180^\circ > \text{RA} > -180^\circ$ and $-60^\circ < \text{dec} < 20^\circ$. To isolate the kSZ signal, an aperture photometry filter with a radius of $2$ arcmin is applied in spherical harmonic space~\citep{Chen2022,Li2024}, adopting a maximum multipole moment of $\ell_{\rm max}=17000$. This filter radius gives the highest S/N of kSZ detection.

The mask map\footnote{\href{https://lambda.gsfc.nasa.gov/product/act/act_dr6.02/act_dr6.02_maps_ancillary_get.html}{srcsamp\_mask.fits}} is used to exclude galaxies located within regions that were applied to high-contrast areas, thereby reducing foreground contamination~\citep{Naess2025}. Additionally, galaxies within approximately $3\sqrt{2}$ arcmin from the edges of either the CMB map or the mask are removed, to minimize edge artifacts introduced by the aperture photometry filter. The resulting sky coverage of the ACT data and its overlapped region with Constant Stellar Mass (CMASS) data is shown in Figure~\ref{fig:sky_distribution}. In this work we only adopt data in overlapped regions to highlight the cosmological benefits of kSZ effects in ideal cases where the galaxy and CMB data are fully overlapped.

\subsection{CMASS}

The CMASS galaxy sample~\citep{Reid2016} is a principal spectroscopic sample from the final Data Release 12 (DR12) of the Baryon Oscillation Spectroscopic Survey (BOSS) -- part of the Sloan Digital Sky Survey III (SDSS-III). CMASS galaxies predominantly reside in massive halos with a mean mass of $2.6 \times 10^{13} \, h^{-1}M_\odot$, a large-scale bias of approximately $2.0$, and a satellite fraction of about $10\%$~\citep{White2011}. These galaxies are characterized by high stellar masses, typically exceeding $10^{11} M_\odot$~\citep{Maraston2009,White2011}, and are largely composed of old stellar populations with low ongoing star formation rates.

We select galaxies within the redshift range $0.43<z<0.75$~\footnote{From \href{https://data.sdss.org/sas/dr12/boss/lss/}{galaxy\_DR12v5\_CMASSLOWZTOT\_North/South.fits.gz} }. After applying the mask, the overlapping area between the ACT footprint and the CMASS sample totals {$4806\ \mathrm{deg}^2$ (Northern Galactic Cap or NGC: $2916\ \mathrm{deg}^2$; Southern Galactic Cap or SGC: $1890 \mathrm{deg}^2$).} The comoving volume of the sample is $2.3\ \mathrm{Gpc}^3/h^3$. The galaxy number density is $1.9\times10^{-4}(h/\rm Mpc)^3$. Using the total weights ($w_{\mathrm{tot}}$) and the Feldman-Kaiser-Peacock (FKP) weights ($w_{\mathrm{FKP}}$)~\citep{Feldman1994,Reid2016,Beutler2017}, the effective redshift of the galaxy sample is determined to be $z_{\mathrm{eff}} = 0.58$. The sky coverage of the CMASS sample is shown in Figure~\ref{fig:sky_distribution}.

\subsection{Mock observations}
\label{subsec:mocks_easy}

In Appendix~\ref{app:model_test}, we conduct in-depth tests to assess the robustness of our joint analysis methodology using mock observations derived from high-resolution dark matter simulations. Specifically, we employ the MultiDark-Patchy mock catalogs~\citep{Kitaura2016} to evaluate the theoretical model of the galaxy power spectrum and the WebSky simulation~\citep{Stein2020} to validate the model of the kSZ power spectrum along with the joint analysis methodology. Further details regarding the adopted mock data are provided in Appendix~\ref{app:mocks_more}.

\section{Methodology}
\label{sec:methodology}

From the aforementioned datasets, we first apply an aperture photometry filter with radius of $2~\rm arcmin$ to detect the kSZ temperature signals at the CMASS galaxy locations. Next, we proceed to measure the multipoles of the galaxy density power spectrum ($\hat{P}_{gg}^{\ell=0,2,4}(k)$) and the density-weighted pairwise kSZ power spectrum ($\hat{P}_{\rm kSZ}^{\ell=1}(k)$). 

The measurement methodology refines the approach of \citet{Li2024} in several key respects. (1) The galaxy density power spectra are measured using \texttt{pypower}\footnote{\url{https://pypower.readthedocs.io}}, an updated version of \texttt{NBODYKIT}~\citep{Hand2018}. This implementation introduces a revised definition of the normalization factor $A$ (differing from the original Equation~(27) in \citealt{Li2024}), which is consistently applied in the estimator for the density-weighted pairwise kSZ power spectrum. (2) Rather than using the effective area of the NGC and SGC, as in Equation~(36) of \citealt{Li2024}, we now employ particle counts from the random catalogs as weights when combining power spectra from both caps. (3) The survey window function effect is incorporated into the theoretical model using the functionality provided by the \texttt{pypower} package~\citep{Beutler2021}. Further methodological details are provided in Appendix~\ref{app:method}.

\section{Theoretical Framework}
\label{sec:theory}
This section outlines the theoretical models employed in our analysis. We begin by presenting the models for the galaxy and kSZ power spectra, then describe the treatment of the the AP effect~\citep{Alcock1979} and finally detail the full set of model parameters to be constrained.

\subsection{Modeling the Power Spectra}
Assuming a uniform average optical depth $\bar{\tau}$ for all galaxies, the kSZ power spectrum can be approximated as
\begin{equation}
\label{eq:PkSZ_to_pv}
    P_{\rm kSZ}(\boldsymbol{k}) \simeq \frac{T_{\rm CMB}\bar{\tau}}{c} P_{\rm pv}(\boldsymbol{k}),
\end{equation}
where $T_{\rm CMB}$ is the CMB temperature and $c$ is the speed of light. The density-weighted pairwise line-of-sight (LOS) velocity power spectrum $P_{\rm pv}$ is related to the galaxy density-momentum cross-power spectrum by~\citep{Sugiyama2018,Li2024}
\begin{equation}
\label{eq:Pgm2Ppv}
    P_{\rm pv}(\boldsymbol{k}) = 2 P_{gp}(\boldsymbol{k}),
\end{equation}
where the subscript $p$ denotes the LOS galaxy momentum field. Substituting Equation~(\ref{eq:Pgm2Ppv}) into Equation~(\ref{eq:PkSZ_to_pv}) yields
\begin{equation}
\label{eq:P_kSZ_ell_2}
    P_{\rm kSZ}(\boldsymbol{k}) \simeq \frac{2T_{\rm CMB}\bar{\tau}}{c} P_{gp}(\boldsymbol{k}).
\end{equation}

The redshift-space galaxy density power spectrum $P_{gg}(k,\mu)$ and the galaxy density-momentum cross-power spectrum $P_{gp}(k,\mu)$ are modeled within the nonlinear perturbation theory framework. We adopt the formulations from~\cite{Vlah2012,Vlah2013,Okumura2014,Saito2014}, as implemented in~\cite{Howlett2019,Qin2019,Qin2025,Qin2025b,Shi2024}:
\begin{equation}
\label{eq:Pgg_Pgm}
    \begin{split}    
    P_{gg}(k,\mu) &= P_{00} + \mu^2(2P_{01} + P_{02} + P_{11}) \\
                  &\quad + \mu^4\left(P_{03} + P_{04} + P_{12} + P_{13} + \frac{1}{4}P_{22}\right)\,, \\
    P_{gp}(k,\mu) &= i\frac{aH}{k}\mu \left[P_{01} + P_{02} + P_{11} \right. \\
                  &\quad \left. + \mu^2\left(\frac{3}{2}P_{03} + 2P_{04} + \frac{3}{2}P_{12} + 2P_{13} + \frac{1}{2}P_{22}\right)\right]\,.
    \end{split}
\end{equation}
Here, $\mu \equiv \cos \theta$ denotes the cosine of the angle between the wavevector $\boldsymbol{k}$ and the LOS direction. Further details of the model calculation are provided in Appendix~\ref{app:theory_calculation}, \LSH{ and the numerical code can be found at \url{https://github.com/shaohongli-code/theoretical_power_spectrum}.}

\subsection{AP Effect}

To incorporate the AP effect -- geometric distortions along and perpendicular to the LOS due to discrepancies between the true and fiducial cosmologies -- we define the scaling factors
\begin{equation}
\label{eq:AP_test}
    \alpha_\parallel = \frac{H^{\rm fid}(z)}{H(z)}, \quad \alpha_\perp = \frac{D_{A}(z)}{D_{A}^{\rm fid}(z)},
\end{equation}
where $H^{\rm fid}(z)$ and $D_{A}^{\rm fid}(z)$ are the fiducial Hubble parameter and angular diameter distance, respectively, evaluated at the effective redshift of the sample. The transformation between the true wavevector components $(k',\mu')$ and the observed values $(k,\mu)$ follows~\citep{Ballinger1996}
\begin{equation}
\label{eq:kmu_AP_test}
    \begin{split}    
    k' &= \frac{k}{\alpha_\perp}\left[1+\mu^2\left(\frac{1}{F^2}-1\right)\right]^{1/2}\, , \\
    \mu' &= \frac{\mu}{F}\left[1+\mu^2\left(\frac{1}{F^2}-1\right)\right]^{-1/2}\, ,
    \end{split}
\end{equation}
with $F \equiv \alpha_\parallel / \alpha_\perp$. The multipoles of the galaxy and kSZ power spectra are then computed via
\begin{equation}
\label{eq:Pgg_PkSZ_AP}
    P_{gg}^{\ell}(k) = \frac{2\ell+1}{2\alpha_\parallel\alpha_\perp^2} \int_{-1}^{1} d\mu\ P_{gg}\left[k'(k,\mu),\mu'(k,\mu)\right]\mathcal{L}_{\ell}(\mu),
\end{equation}
\begin{equation}
    P_{\rm kSZ}^{\ell}(k) = \frac{2\ell+1}{2\alpha_\parallel\alpha_\perp^2} \int_{-1}^{1} d\mu\ P_{\rm kSZ}\left[k'(k,\mu),\mu'(k,\mu)\right]\mathcal{L}_{\ell}(\mu).
\end{equation}

\subsection{Model Parameters}

Our analysis follows the classic GC analysis methodology, aiming to constrain two sets of parameters. First, we constrain a set of cosmological observables, including the linear growth rate $f$ and the AP scaling parameters $\alpha_{\parallel}$ and $\alpha_\perp$. Subsequently, we replace these observables with cosmological parameters $H_0$, $\Omega_m$, and $w_0$ using the relations detailed in Appendix~\ref{app:obs-to-cos} and directly fit these cosmological parameters. In both analyses, the linear matter power spectrum is fixed using the best-fit cosmological parameters from Planck18~\citep{PLANK2018}. We do not expect this choice to bias our results, because the large-scale information of $f$ is primarily derived from ratios between power spectrum multipoles, where the linear power spectrum cancels out. Moreover, the AP parameters are constrained by the isotropy of the galaxy distribution and are independent of the shape of the power spectrum. We also fit for the mean optical depth $\bar{\tau}$ on both stages -- a parameter that encapsulates information on the gas density distribution within and around dark matter halos~\citep[e.g.,][]{ZhengY2024}.

To account for galaxy bias and nonlinear RSD effects, such as the Fingers-of-God (FoG) suppression~\citep{Jackson1972}, we include several nuisance parameters: the linear bias $b_1$, the second-order bias $b_2$, two velocity dispersion parameters, $\sigma^2_{v,1}$ and $\sigma^2_{v,2}$, which improve the model accuracy at nonlinear scales beyond that of the single $\sigma_v^2$ model~\citep{Vlah2012,Howlett2019}, and a residual shot-noise parameter $N_{\rm sn}$, which addresses potential imperfections in the subtraction of the shot-noise term. The complete set of free parameters in our model and the flat priors adopted in the {likelihood} analysis are summarized in Table~\ref{tab:prior}. No CMB priors are used during the fitting.

\begin{table}[t!]
\caption{\label{tab:prior}Uniform priors of free parameters.}
\begin{ruledtabular}
\begin{tabular}{cc|cc}

Cosmological &Prior&Cosmological &Prior \\
Observable & &Parameter &\\
\hline
$f$ & [0., 2.] &$H_0$ & [50,100] \\
$\alpha_\parallel$ & [0.5,1.5] & $\Omega_{\rm m}$ & [0.,1.] \\
$\alpha_\perp$ & [0.5,1.5] & $w_0$ & [-3.,1.] \\
\hline
Astrophysical&&&\\
Parameter &&&\\
 ${\rm lg}\ \bar{\tau}$ & [-6.,0.]
\end{tabular}
\begin{tabular}{cccc}
Nuisance &Prior&Nuisance &Prior \\
 Parameter&& Parameter& \\
$b_1$ & [0,5] & $b_2$ & [-10,10] \\
$\sigma^2_{v,1}$ & [0,200] & $\sigma^2_{v,2}$ & [0,200] \\
$N_{\rm sn}$ & [$-10^4$,$10^4$] &   & \\
\end{tabular}
\end{ruledtabular}
\end{table}

\section{Results}
\label{sec:results}

In this section, we present constraints on the model parameters from a joint Markov Chain Monte Carlo (MCMC) analysis of the galaxy power spectrum multipoles $\hat{P}_{gg}^{\ell=0,2,4}(k)$ and the kSZ dipole $\hat{P}_{\rm kSZ}^{\ell=1}(k)$, which are shown in the left panels of Figure~\ref{fig:CMASS}, assuming a Planck18 $\Lambda$CDM fiducial cosmology~\citep{PLANK2018} with $\Omega_m = 0.31$, $\Omega_b h^2 = 0.02242$, $h = 0.6766$, $\sigma_8 = 0.8102$, $n_s = 0.9665$, and $\sum m_\nu = 0.06$ eV. The analysis is performed in the wavenumber range $k \sim [0.01, 0.15]\,h\,\mathrm{Mpc}^{-1}$, which was validated in Appendix~\ref{app:model_test} to reliably recover the input cosmology from mock data. 

\subsection{Constraints on cosmological observables}
\begin{figure*}[t!]
\includegraphics[width=\columnwidth]{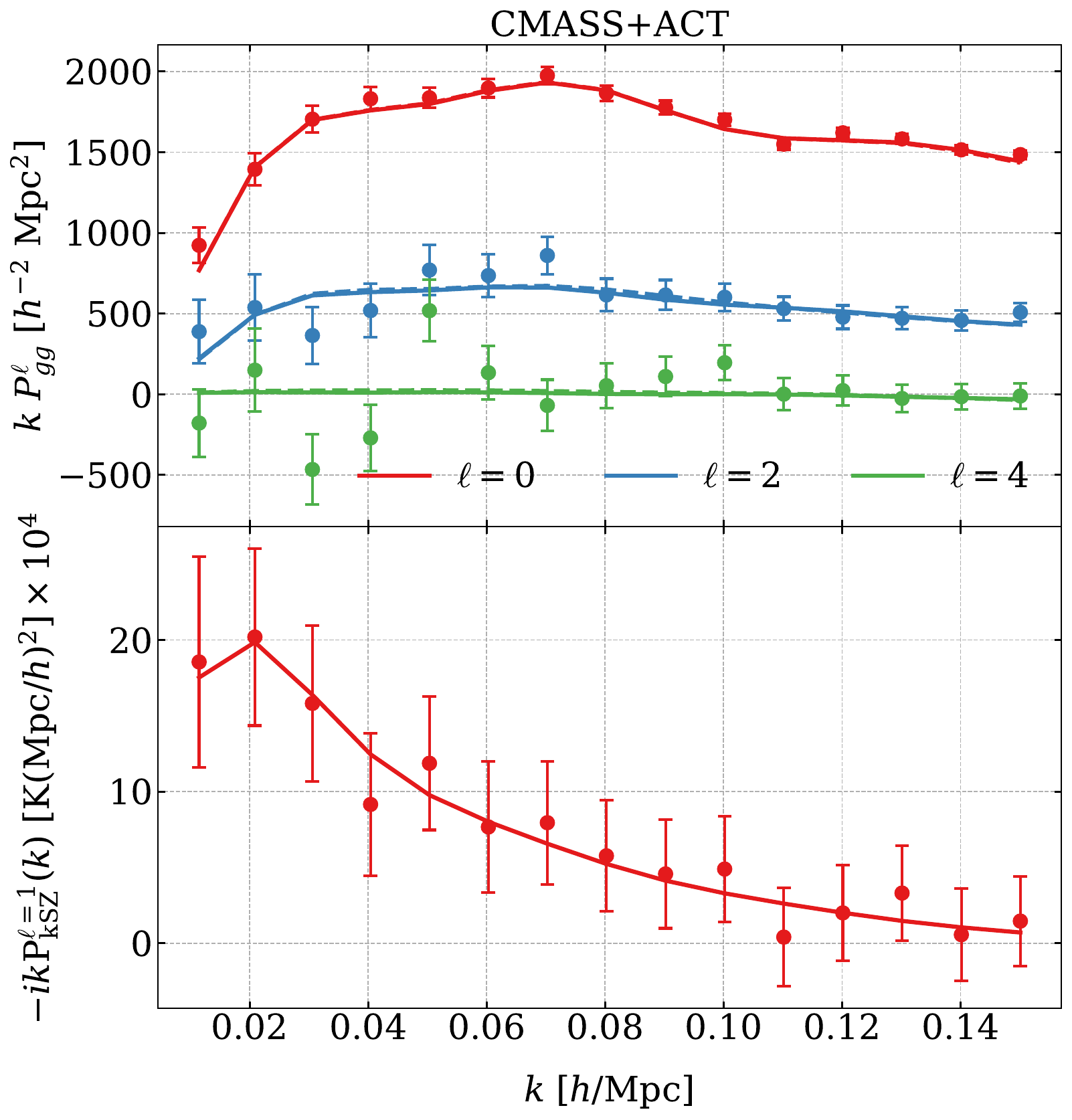}
\includegraphics[width=1.1\columnwidth]{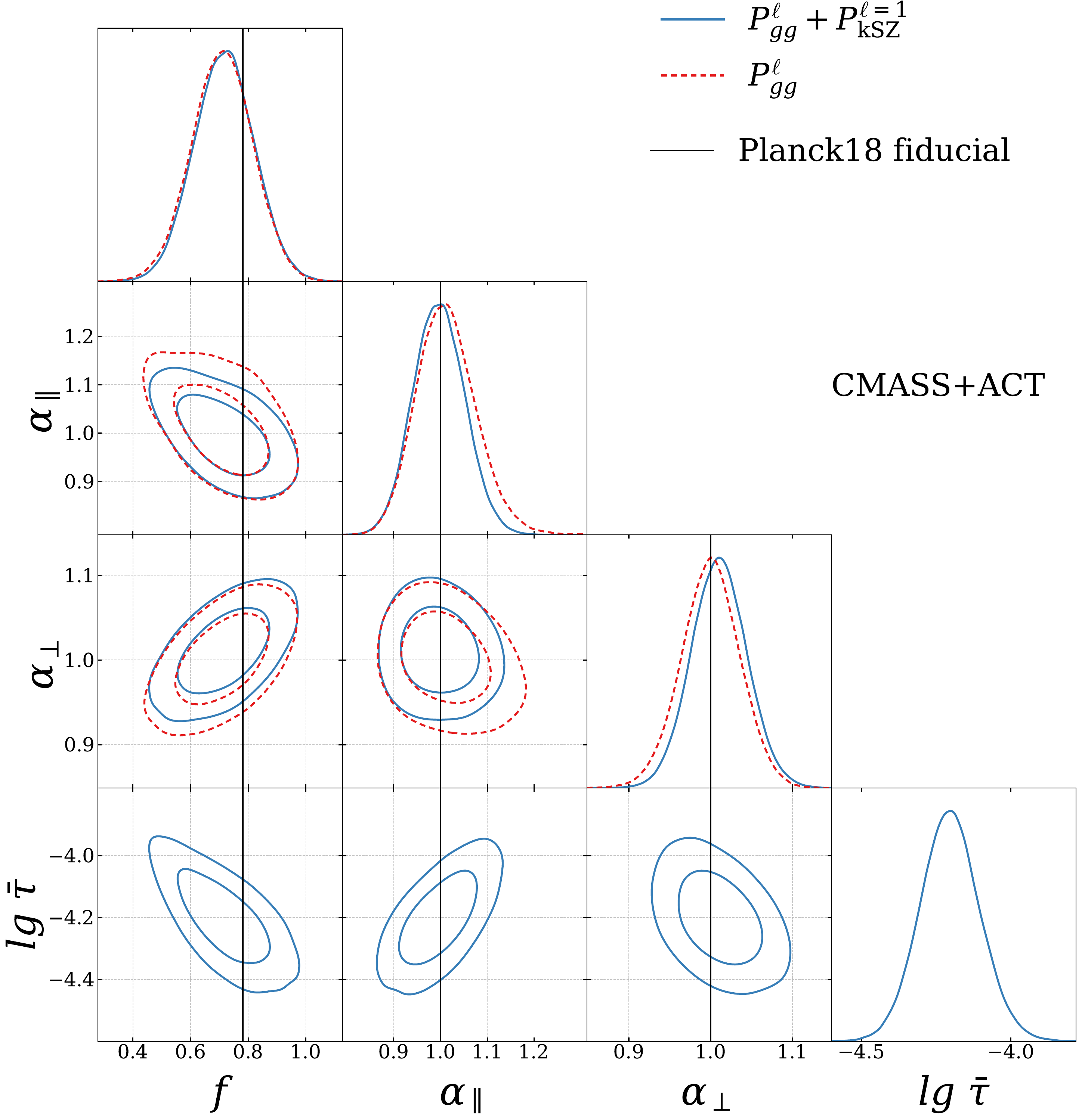}
\caption{\label{fig:CMASS} Results of CMASS + ACT analysis.
\textit{Upper left:} multipoles of the galaxy density power spectrum. The dashed lines indicate the best-fit model by fitting the galaxy multipoles alone, while solid lines show the results from the joint analysis.  \textit{Lower left:} the kSZ power spectrum dipole along with the best-fitted model from the joint analysis (solid line). The covariance matrices of these power spectra are computed using a jackknife resampling method. The S/N of this kSZ dipole is estimated to be $\sim$7, as detailed in Appendix~\ref{app:method}.
\textit{Right:} posterior distributions of the cosmological observables. Blue solid contours correspond to results of the joint analysis, and red contours represent the constraints from galaxy multipoles only. The black vertical lines mark the fiducial values $f=\Omega^{\rm fid}_{\rm m}(z_{\rm eff})^{0.55}$, $\alpha_\parallel=1$ and $\alpha_\perp=1$, where $\Omega^{\rm fid}_{\rm m}(z_{\rm eff})$ is the matter density at the effective redshift $z_{\rm eff}$, based on the fiducial cosmology.}
\end{figure*}

The right panel of Figure~\ref{fig:CMASS} presents the constraints on the key cosmological observables. After marginalizing over all nuisance parameters, the joint GC+kSZ analysis yields:

\LSH{
{
\begin{itemize}
  \item $f = 0.717^{+0.104}_{-0.105}$;
  \item $\alpha_\parallel = 0.997^{+0.055}_{-0.054}$; and
  \item $\alpha_\perp = 1.010^{+0.034}_{-0.035}$.
\end{itemize}
}
}
These constraints are tighter than those from the GC-only analysis {({$f = 0.711^{+0.105}_{-0.106}$, $\alpha_\parallel = 1.011^{+0.065}_{-0.060}$, $\alpha_\perp = 1.00^{+0.035}_{-0.036}$}), demonstrating the added value of the kSZ effect. All measured values are consistent with the fiducial cosmology ($f=0.782$, $\alpha_\parallel=1.0$, $\alpha_\perp=1.0$)} within $1\sigma$ uncertainties and show agreement with the \textit{Planck} 2018 prediction within the $\Lambda$CDM framework.

To quantitatively assess the enhancement in constraining power from the joint analysis, we compute the Figure of Merit (FoM) for pairs of parameters, defined as the inverse of the area enclosed by their $1\sigma$ confidence contour $A_{1\sigma}$:
\begin{equation}
\label{eq:FoM}
    \mathrm{FoM} = \frac{1}{A_{1\sigma}}\,,
\end{equation}
where a larger FoM corresponds to a tighter constraint.

For the CMASS+ACT data, the FoM improves by:
\LSH{
{
\begin{itemize}
    \item 15.6\% for the $f$--$\alpha_\parallel$ pair;
    \item 7.4\% for $f$--$\alpha_\perp$; and
    \item 15.3\% for $\alpha_\parallel$--$\alpha_\perp$.
\end{itemize}
}
}
These positive improvements across all parameter pairs confirm that the inclusion of the kSZ effect considerably strengthens the cosmological constraints.

\subsection{Constraints on cosmological parameters}

\begin{figure}[t!]
\includegraphics[width=\columnwidth]{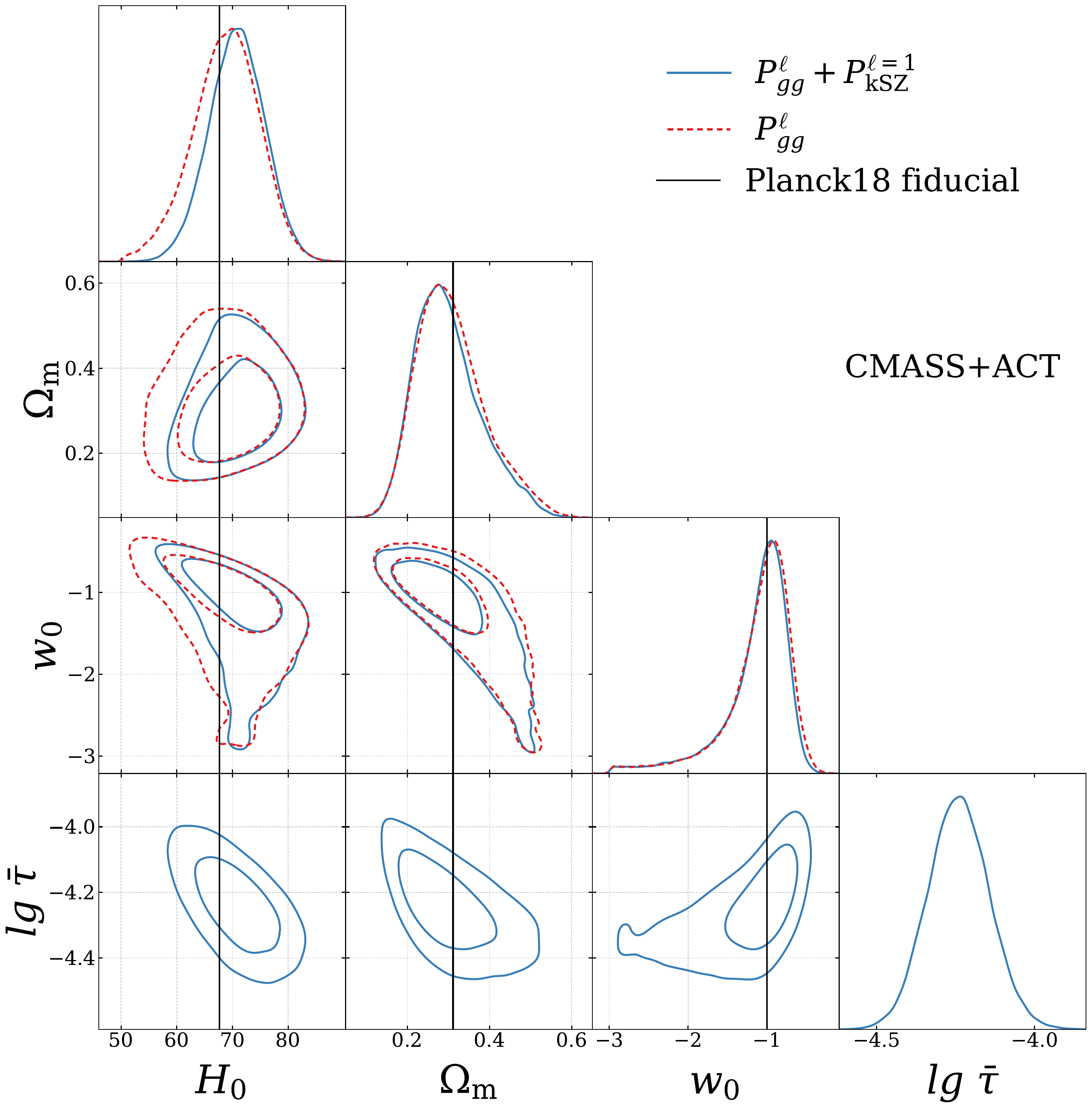}
\caption{\label{fig:CMASS_H0Omw0} Similar to the right panel of Figure~\ref{fig:CMASS}, but for the cosmological parameters $H_0$, $\Omega_{\rm m}$ and $w_0$.}
\end{figure}

We now present direct constraints on the fundamental cosmological parameters, with the results displayed in Figure~\ref{fig:CMASS_H0Omw0}. After marginalizing over all nuisance parameters, the joint GC+kSZ analysis yields
\LSH{
{
\begin{itemize}
    \item $H_0 = 70.82^{+4.94}_{-5.01}$;
    \item $\Omega_{\rm m} = 0.290^{+0.092}_{-0.068}$; and
    \item $w_0 = -1.038^{+0.245}_{-0.437}$.
\end{itemize}
}
}
For the GC-only analysis, we find {{$H_0 = 69.02^{+5.65}_{-6.14}$, $\Omega_{\rm m} = 0.298^{+0.093}_{-0.072}$ and $w_0 = -1.014^{+0.256}_{-0.431}$}}. All fiducial values lie within the $1\sigma$ uncertainties of our measurements. 

The joint analysis improves the FoM by:
\LSH{
{
\begin{itemize}
    \item 20.5\% for the $H_0$--$\Omega_{\rm m}$ pair;
    \item 19.7\% for $H_0$--$w_0$; and
    \item 10.0\% for $\Omega_{\rm m}$--$w_0$.
\end{itemize}
}
}
This represents a stronger enhancement (averaging $\sim$\LSH{20\%}) compared to the $\sim$\LSH{15\%} improvement seen for the cosmological observables ($f$, $\alpha_\parallel$, $\alpha_\perp$), likely due to the kSZ effect breaking more degeneracies in a complementary direction within this different parameter space. In Appendix~\ref{app:fisher_matrix}, we use the Fisher matrix to predict the improvement of FoM, and the similar stronger enhancement is found.

Furthermore, we constrain the mean optical depth of the galaxy sample to \LSH{$\lg\bar{\tau} = -4.20 \pm 0.10$ in Figure~\ref{fig:CMASS} and $\lg\bar{\tau} = -4.24 \pm 0.10$} in Figure~\ref{fig:CMASS_H0Omw0}. This quantifies the integrated column density of free electrons within the 2 arcmin AP filter around galaxies. This result serves as a demonstration the joint constraining of cosmological and astrophysical parameters in a combined analysis of GC and the kSZ effect, which is a methodology we expect to become standard with future, higher-quality data.

We find that the measured $\bar{\tau}$ for CMASS is lower than that derived from the WebSky-CMASS mock catalogs (Figure~\ref{fig:websky_CMASS}). This discrepancy can be attributed to at least two factors. (1) A positive correlation exists between halo mass and optical depth~\citep[e.g.,][]{Chen2022,Li2024}. As shown in Figure~\ref{fig:mass_distribution_websky_CMASS} of Appendix~\ref{app:mass_dis_websky}, the halo mass distribution in our WebSky-CMASS mock catalog is skewed toward higher masses compared to the observationally inferred CMASS halo mass distribution from~\citet{Schaan2021}. The mean halo mass in WebSky-CMASS is {$4.9 \times 10^{13} \, h^{-1} M_\odot$}, nearly twice the value of $2.6 \times 10^{13} \, h^{-1} M_\odot$ estimated for CMASS~\citep{White2011}. (2) Approximately 10\% of CMASS galaxies are satellites~\citep{White2011}. The associated miscentering of these satellites with respect to their dark matter halo centers can significantly dilute the observed kSZ signal~\citep{Hadzhiyska2023}.

{The posterior distributions of all parameters from CMASS+ACT data are presented in Figures~\ref{fig:corner_CMASS_all_parameters} and~\ref{fig:corner_CMASS_all_parameters_H0Omw0}.}

\section{Conclusion}
\label{sec:conclusion}

We have presented in this work the first joint cosmological analysis of the GC and kSZ effects using real observational data, simultaneously constraining both cosmological and astrophysical parameters. By combining the galaxy density power spectrum multipoles $\hat{P}_{gg}^{\ell=0,2,4}(k)$ with the pairwise kSZ power spectrum dipole $\hat{P}_{\rm kSZ}^{\ell=1}(k)$, we have established and implemented a robust multitracer methodology that significantly enhances parameter constraints.

Our analysis of the CMASS galaxy sample from BOSS and the ACT DR6 CMB map yields consistent constraints on key cosmological quantities. For the cosmic growth rate and expansion history, {the joint analysis gives \LSH{$f = 0.717^{+0.104}_{-0.105}$, $\alpha_\parallel = 0.997^{+0.055}_{-0.054}$, and $\alpha_\perp = 1.010^{+0.034}_{-0.035}$}, with the FoM improving by approximately \LSH{12\%} across different parameter pairs compared to GC-only constraints.} For the cosmological parameters, we obtain \LSH{{$H_0 = 70.82^{+4.94}_{-5.01}$, $\Omega_{\rm m} = 0.290^{+0.092}_{-0.068}$, and $w_0 = -1.038^{+0.245}_{-0.437}$}}, with an average FoM improvement of approximately \LSH{17\%}. This substantial enhancement demonstrates that the kSZ effect provides independent cosmological information that effectively breaks degeneracies in parameter space. Additionally, we constrain the mean optical depth of the galaxy sample to \LSH{$\lg\bar{\tau} = -4.24 \pm 0.10$}, showcasing the ability to simultaneously probe astrophysical properties.

Looking forward, this joint analysis framework presents a powerful approach for extracting cosmological and astrophysical information from upcoming spectroscopic galaxy and CMB surveys. With future data from DESI~\citep{DESI2016}, PFS~\citep{Takada2014}, Euclid~\citep{Euclid2020}, CSST~\citep{CSST2025}, and SO~\citep{Simons2019}, CMB-S4~\citep{CMBS42019-2}, complemented by refined theoretical models, such as the effective field theory (EFT) of large-scale structure via which we can conduct full-shape analysis~\citep[e.g.,][]{ChenSF2025}, we anticipate achieving unprecedented precision in constraining both cosmological parameters and the astrophysics of baryonic processes in galaxies and halos~\citep{LiZY2025}.

\begin{acknowledgements}
{\it Acknowledgement:} \ZY{We thank Fei Qin for useful discussions. We thank Naonori Sugiyama for providing valuable comments and suggestions that helped us significantly improve the manuscript quality.} Y.Z. acknowledges the support from the National SKA Program of China (2025SKA0150104) and the National Natural Science Foundation of China (NFSC) through grant 12203107.
\end{acknowledgements}

\appendix

\section{Relations between cosmological observables and parameters}
\label{app:obs-to-cos}

{This appendix presents the relations between the cosmological observables ($f$, $\alpha_\parallel$ and $\alpha_\perp$) and the cosmological parameters ($H_0$, $\Omega_{\rm m}$ and $w_0$). We assume a flat Universe. The expansion rate of the Universe at the redshift z is described by the Hubble parameter $H(z)=H_0 E(z)$, where $H_0$ is the present-day value of $H(z)$ and the time-dependent function $E(z)$ is expressed as}
\begin{equation}
\label{eq:E_z}
    \begin{split}    
    E^2(z)=\Omega_{\rm m}(1+z)^3+\Omega_{\rm DE}(1+z)^{3(1+w_0)}.
    \end{split}
\end{equation}
Here, $\Omega_{\rm m}$ and $\Omega_{\rm DE}$ are the present-day energy density fractions of matter and dark energy, respectively, with $\Omega_{\rm m}+\Omega_{\rm DE}=1$. When the equation-of-state parameter for dark energy, denoted by $w_0$, is not equal to $-1$, the assumed Universe model deviates from the standard cosmological model.

The angular diameter distance is $D_A(z)=(1+z)^{-1}\chi(z)$ with the comoving distance
\begin{equation}
\label{eq:chi}
    \begin{split}    
    \chi(z)=\int^z_0\frac{c}{H(z')}dz'.
    \end{split}
\end{equation}

The growth rate $f$ can be parameterized as
\begin{equation}
\label{eq:f_z}
    \begin{split}    
    f(z)=[\Omega_{\rm m}(z)]^\gamma\, ,
    \end{split}
\end{equation}
where $\Omega_{\rm m}(z)=\Omega_{\rm m}(1+z)^3/E^2(z)$ is the time-dependent matter density and the index $\gamma$ specifies a model of gravity. In this Letter, we adopt $\gamma=0.55$ which satisfies general relativity~\citep{Peebles1980,Linder2005}. By fixing $\gamma$, we effectively use all information from $f$ to constrain the cosmic expansion history as well.

\section{Mock observations}
\label{app:mocks_more}

In order to verify the accuracy and precision of the theoretical models, we introduce in this section two sets of simulation catalogs that are similar to the observation data. In particular, one set of them, from the WebSky simulation, is used to investigate the robustness of our joint analysis methodology. 

\subsection{CMASS mocks}
We use the MultiDark-Patchy mock catalogs~\citep{Kitaura2016} to test the theoretical model of the galaxy power spectrum. These mock catalogs were constructed to enable a reliable analysis of baryon acoustic oscillation (BAO) and RSD in the final dataset of BOSS, including CMASS. There are 2048 mock samples each for the NGC and the SGC. 
{The veto masks are used, and the masks used for these mocks are the same as CMASS. The fiducial cosmological parameters of these mock catalogs are: $\Omega_m = 0.3071$, $\Omega_b h^2 = 0.02214$, $h = 0.6777$, $\sigma_8 = 0.8288$, $n_s = 0.9611$, and $\sum m_\nu = 0.06$ eV.} The test results of the mock catalogs are presented in Appendix~\ref{subapp:cmassmock_test}.

\subsection{WebSky simulation}
We employ the WebSky simulation~\citep{Stein2020} to validate the joint analysis methodology. WebSky is a widely used suite of high-fidelity simulated sky maps that incorporate multiple cosmological signals, such as the primary cosmic microwave background (CMB), the thermal  Sunyaev–Zel'dovich effects and kSZ effects, the cosmic infrared background (CIB), and radio sources. These simulations are generated using a fast, self-consistent approach based on halo light-cones constructed from a large N-body simulation. From the WebSky simulation, we construct two distinct halo samples. The first, referred to as WebSky-CMASS, is designed to match both the sky coverage and the redshift distribution of the observational CMASS sample, achieved by selecting the most massive halos. The second sample, termed WebSky-allsky, covers the full sky while maintaining the same comoving number density as the CMASS sample in redshift. This all-sky sample serves as a reference for evaluating the accuracy of power spectrum models with the impact of cosmic variance highly suppressed.

We generate the corresponding simulated CMB map by combining multiple microwave components from the WebSky simulation to replicate the ACT DR6 f150 map characteristics. The map is constructed at a HEALPix resolution of $N_{\rm side}=4096$ and incorporates the following components: kSZ, tSZ at 150 GHz, CIB at 145 GHz, lensed CMB, and instrumental noise with a level of $14~\mu\text{K} \cdot \text{arcmin}$. The composite map is then smoothed with a Gaussian beam of $\text{FWHM} = 1.42'$. The cosmological parameters adopted in the WebSky simulation follow the fiducial values described in \cite{Stein2020}. The test results based on the WebSky samples are provided in Appendices~\ref{subapp:webskycmass_test} and~\ref{subapp:webskyallsky_test}.

\section{Details of the power spectrum estimators}
\label{app:method}
The estimators for the multipoles of the galaxy density power spectrum and the pairwise kSZ power spectrum are constructed as follows \citep{Feldman1994,Yamamoto05,Hand2017,Sugiyama2018}:
\begin{equation}
\label{eq:P_AP_test}
    \begin{split}    
    \hat{P}^{\ell}_{gg}(k)&=\frac{2\ell+1}{A}\int\frac{d\Omega_k}{4\pi}[\delta n (\boldsymbol{k})\delta n ^* _\ell(\boldsymbol{k})-P_\ell^{\rm noise}(\boldsymbol{k})]\,, \\
    \hat{P}^{\ell}_{\rm kSZ}(k)&=-\frac{2\ell+1}{A}\int\frac{d\Omega_k}{4\pi}[\delta T (\boldsymbol{k})\delta n ^* _\ell(\boldsymbol{k})-\delta T ^* (\boldsymbol{k})\delta n _\ell(\boldsymbol{k})]\,,
    \end{split}
\end{equation}
with
\begin{equation}
   \begin{split} 
        \delta n _\ell(\boldsymbol{k})&=\int d^3 s e^{-i\boldsymbol{k}\cdot{} \boldsymbol{s}}w(\boldsymbol{s})[n_g (\boldsymbol{s})-\alpha n_r (\boldsymbol{s})]\mathcal{L}_\ell(\hat{\boldsymbol{k}}\cdot{}\hat{\boldsymbol{s}})\,, \\
        \delta T(\boldsymbol{k})&=\int d^3 s e^{-i\boldsymbol{k}\cdot{} \boldsymbol{s}}w(\boldsymbol{s})\delta T(\boldsymbol{s})\,.
   \end{split}
\end{equation}
Here, $n_g (\boldsymbol{s})$ and $n_r (\boldsymbol{s})$ denote the number densities of the galaxy catalog and the random catalog, respectively. The random catalog density reflects the expected mean galaxy density and incorporates the survey geometry, including the angular mask and radial selection function. The weight function is defined as $w(\boldsymbol{s})=w_{\rm tot}\cdot w_{\rm FKP}$, where $w_{\rm tot}$ corrects for observational systematics to better approximate the true galaxy density field~\citep{Reid2016}, and $w_{\rm FKP}$ optimizes the signal-to-noise ratio in power spectrum estimation~\citep{Feldman1994}. The factor $\alpha$ normalizes the random catalog to match that of the galaxy catalog density. The shot-noise term $P_\ell^{\rm noise}(\boldsymbol{k})$ and the normalization factor $A$ are given by
\begin{equation}
   \begin{split} 
        P_\ell^{\rm noise}(\boldsymbol{k})&=(1+\alpha)\int d^3 s e^{-i\boldsymbol{k}\cdot{} \boldsymbol{s}}\bar{n}_g (\boldsymbol{s})w^2(\boldsymbol{s})\mathcal{L}_\ell(\hat{\boldsymbol{k}}\cdot{}\hat{\boldsymbol{s}})\,, \\
        A&=\int d^3 s \bar{n}_g^2 (\boldsymbol{s})w^2(\boldsymbol{s})\,.
   \end{split}
\end{equation}

Furthermore, $\delta T(\boldsymbol{s})$ denotes the kSZ temperature fluctuation field, which is constructed using galaxy tracers and extracted from the CMB map via aperture photometry filtering. The filter is applied with an inner radius of $2\ \rm arcmin$ and implemented in spherical harmonic space. For the WebSky simulation, the filtering is performed using the \texttt{healpy}\footnote{\href{https://healpy.readthedocs.io/en/latest/index.html}{https://healpy.readthedocs.io/en/latest/index.html}} package, while for ACT data the \texttt{pixell}\footnote{\href{https://pixell.readthedocs.io/en/latest/readme.html}{https://pixell.readthedocs.io/en/latest/readme.html}} library is employed. The weighted kSZ temperature field is subtracted by its redshift-dependent mean, where the averaging is performed using a Gaussian weight with a standard deviation of 0.01. Further details can be found in~\cite{Li2024}.

The galaxy power spectrum $\hat{P}^{\ell}_{gg}(k)$ is estimated using \texttt{pypower}\footnote{\href{https://pypower.readthedocs.io}{https://pypower.readthedocs.io}}, a modified version of \texttt{NBODYKIT}~\citep{Hand2018} that incorporates an improved numerical method for computing the normalization factor $A$. The estimation of the kSZ power spectrum $\hat{P}^{\ell}_{\rm kSZ}(k)$ follows the methodology described in~\cite{Li2024}, except for the treatment of the normalization $A$. To discretize the galaxy distribution and kSZ temperature field, we employ the triangular-shaped cloud (TSC) scheme for grid assignment and apply interlacing technique to reduce numerical artifacts such as aliasing and window function effects introduced during gridding. The power spectrum is computed in a periodic cubic grid of size $512^3$, with box side lengths of $(1700, 3350, 850)\ {\rm Mpc}/h$ for NGC and $(1100, 2600, 1100)\ {\rm Mpc}/h$ for SGC. For the full-sky case, a cubic box of side length $3700\ {\rm Mpc}/h$ is used. 
The final power spectrum measurements and effective redshift are derived as weighted averages: for the CMASS data, weights are given by the number of galaxies weighted by the product $w_{\mathrm{tot,i}}\times w_{\mathrm{FKP,i}}$ over galaxies in the NGC and SGC regions; for mock catalogs, weights are given by the number of random points weighted by $w_{\mathrm{FKP,i}}$ in the NGC and SGC parts of random catalogs. Here, the subscript $i$ in $w_{\mathrm{tot,i}}$ and $w_{\mathrm{FKP,i}}$ denotes the $i$-th galaxy or random point.

Using the \texttt{pypower} package~\citep{Beutler2021}, we compute the window function matrices from the random catalog. Both the window function effect and the wide-angle effect~\citep[e.g.,][]{Beutler2019, Reimberg2016, Castorina2018} are incorporated into the theoretical models in Fourier space.

The covariance matrices for both the CMASS data and the WebSky simulation are computed using a resampling approach based on the delete-one jackknife (JK) method~\citep{Sugiyama2018,Li2024}. The sky is partitioned into {1024} subregions via the {\it kmeans} algorithm\footnote{\href{https://github.com/esheldon/kmeans_radec/}{https://github.com/esheldon/kmeans\_radec/}} applied to the random catalog to generate subsamples. The robustness of this choice is validated in Appendix~\ref{app:cov_validation}. For CMASS mock catalogs, the covariance matrix is estimated directly from 2048 MultiDark-Patchy mocks~\citep{Beutler2017}. Finally, all inverse covariance matrices are corrected using the Hartlap factor~\citep{Hartlap2007} to account for statistical bias. {In table~\ref{tab:Hartlap}, we list Hartlap factors of typical subsample numbers adopted to estimate the covariance, for both the GC-only analysis and the joint analysis.}

\begin{table}[t!]
\caption{\label{tab:Hartlap} {Hartlap correction factors.}}
\centering
\begin{tabular}{c|c|c}
\hline
Subsample Number & Hartlap (GC-only) & Hartlap (Joint) \\
\hline
1024 & 0.955 & 0.940  \\
2048 & 0.978 & 0.970  \\
4096 & 0.989 & 0.985  \\
\hline
\end{tabular}
\end{table}

To estimate the S/N of the kSZ dipole measurement, we model the power spectrum as a linear function with a single amplitude parameter $\mathcal{A}$, such that $\bar{P}^\ell_{\rm kSZ} = \mathcal{A} P^\ell_{\rm kSZ}$. Here, $P^\ell_{\rm kSZ}$ is computed using Equation~(\ref{eq:P_kSZ_ell_2}), with all model parameters fixed to the best-fit values from the GC-only analysis. The corresponding $\chi^2$ statistic is given by
\begin{equation}
\label{eq:chi2}
    \chi^2(\mathcal{A}) = \left[\hat{P}^{\ell}_{\rm kSZ} - \bar{P}^{\ell}_{\rm kSZ}(\mathcal{A})\right]^{\rm T} \hat{C}^{-1} \left[\hat{P}^{\ell}_{\rm kSZ} - \bar{P}^{\ell}_{\rm kSZ}(\mathcal{A})\right]\,,
\end{equation}
where $\hat{C}^{-1}$ is the precision matrix, and $\hat{P}^{\ell}_{\rm kSZ}$ is the measured kSZ dipole. Then the S/N is estimated as
\begin{equation}
\label{eq:SN}
    \frac{S}{N} = \sqrt{\chi^2_{\rm null} - \chi^2_{\rm min}}\,,
\end{equation}
where $\chi^2_{\rm null} = \chi^2(\mathcal{A}=0)$ and $\chi^2_{\rm min} = \chi^2(\mathcal{A} = \mathcal{A}_{\rm bestfit})$. For this linear single-parameter model, the best-fit amplitude can be derived analytically as
\begin{equation}
\label{eq:tau}
    \begin{aligned}
        \mathcal{A}_{\rm bestfit}=\frac{(\hat{P}^{\ell}_{\rm kSZ})^{\rm T} \hat{C}^{-1}\bar{P}^{\ell}_{\rm kSZ}(\mathcal{A}=1)}{[P^{\ell}_{\rm kSZ}(\mathcal{A}=1)]^{\rm T}\hat{C}^{-1}\bar{P}^{\ell}_{\rm kSZ}(\mathcal{A}=1)}\,.
    \end{aligned}
\end{equation}
The resulting S/N for our measurement is {7.2}.

\section{Model Validation}
\label{app:model_test}

In this section, we validate our pipeline for the GC-only analysis using the CMASS mock catalogs. The robustness of the joint analysis and its superiority over the GC-only approach are further demonstrated with the WebSky simulation.

\subsection{Tests on CMASS mock}
\label{subapp:cmassmock_test}
\begin{figure*}
\includegraphics[width=0.48\textwidth]{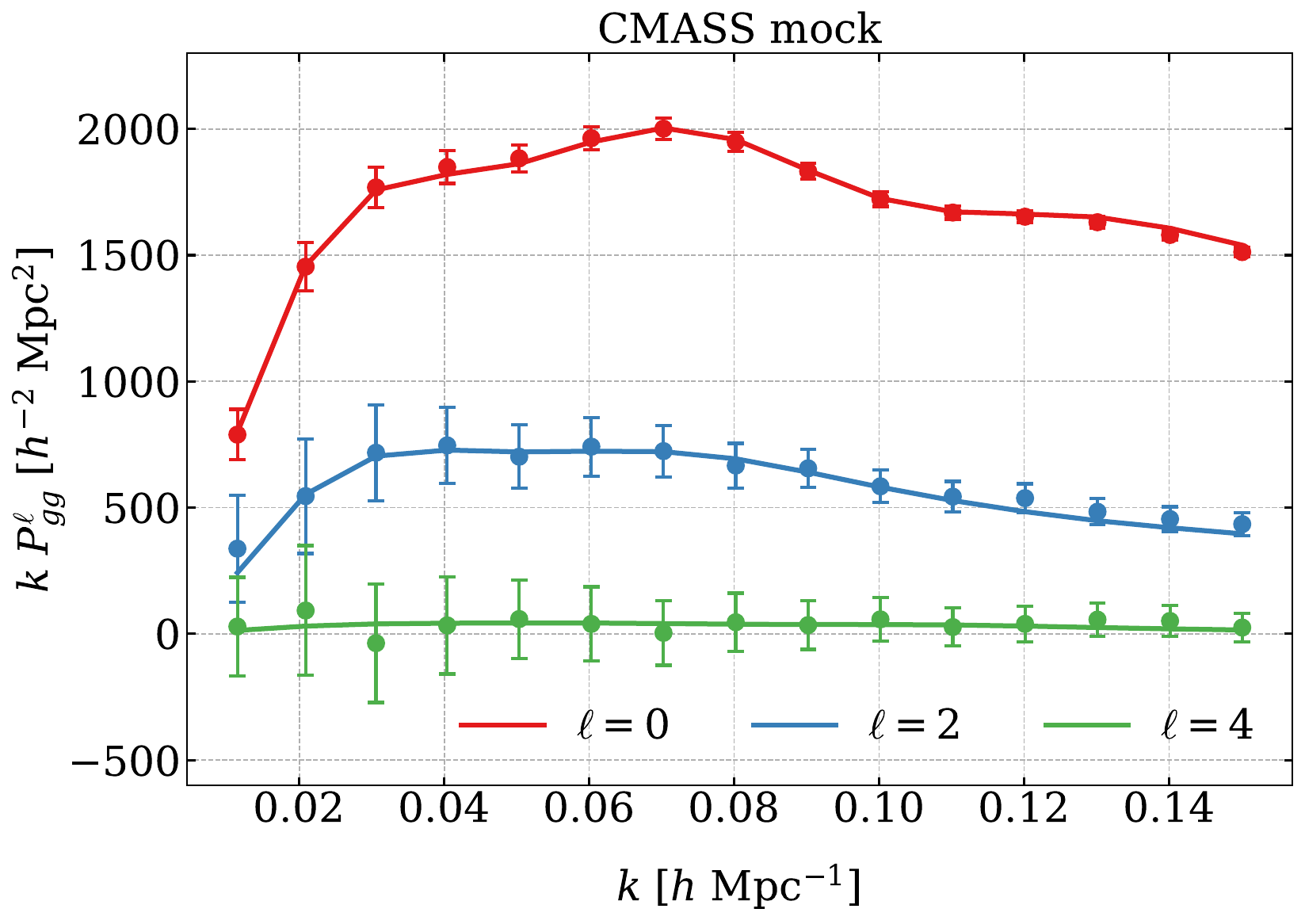}
\includegraphics[width=0.5\textwidth]{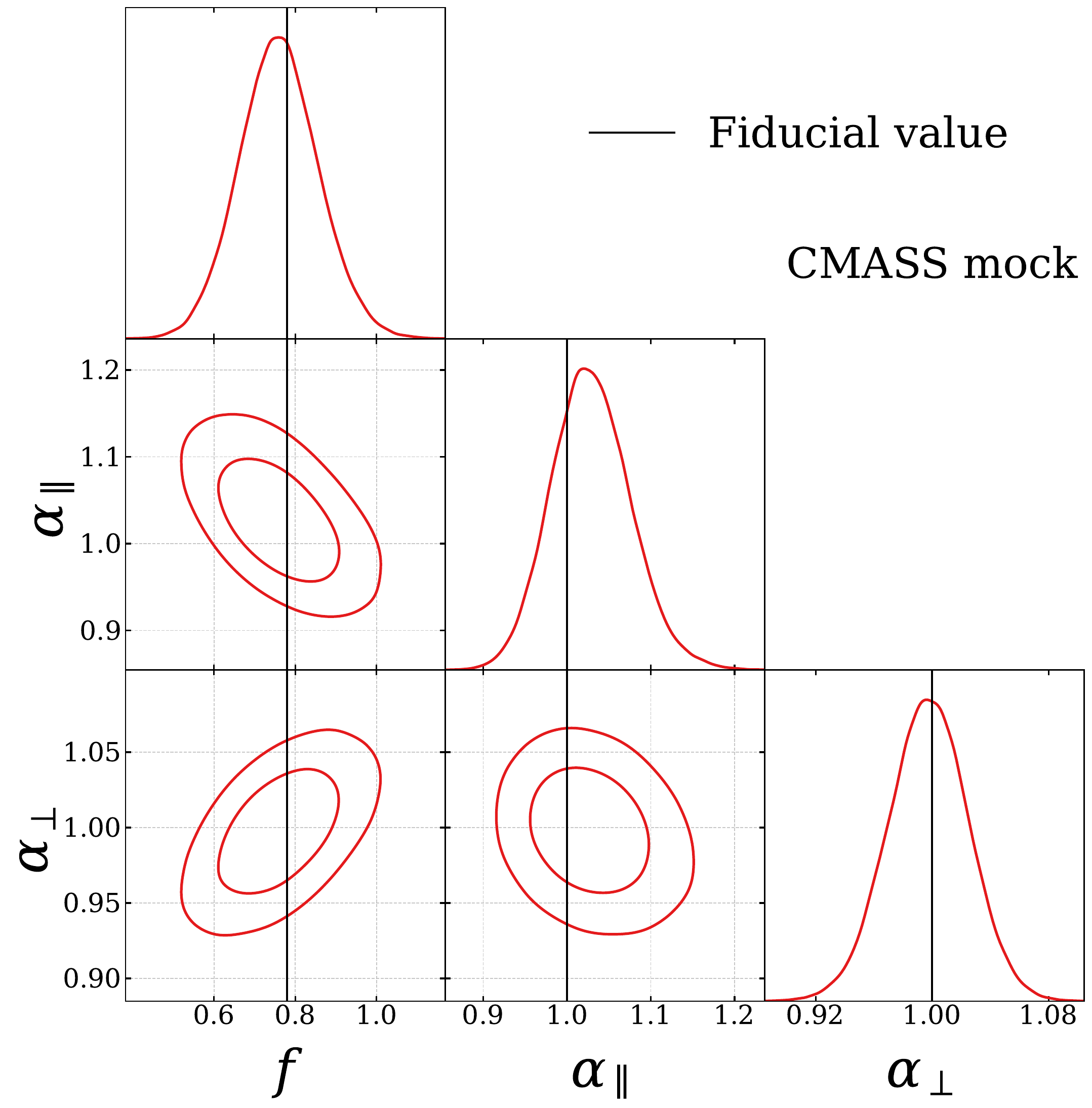}
\caption{\label{fig:CMASS_mock} \textit{Left:} the multipoles of the CMASS mock galaxy power spectra. The center points with error bars represent the average values of the measurements of the multipoles. The solid lines are the best-fit multipoles. \textit{Right:} posterior distributions of three free cosmological parameters -- $f$, $\alpha_\parallel$ and $\alpha_\perp$. The black solid lines represent the theoretical values of fiducial cosmology.}
\end{figure*}
\begin{figure*}
\centering
\includegraphics[width=0.45\textwidth]{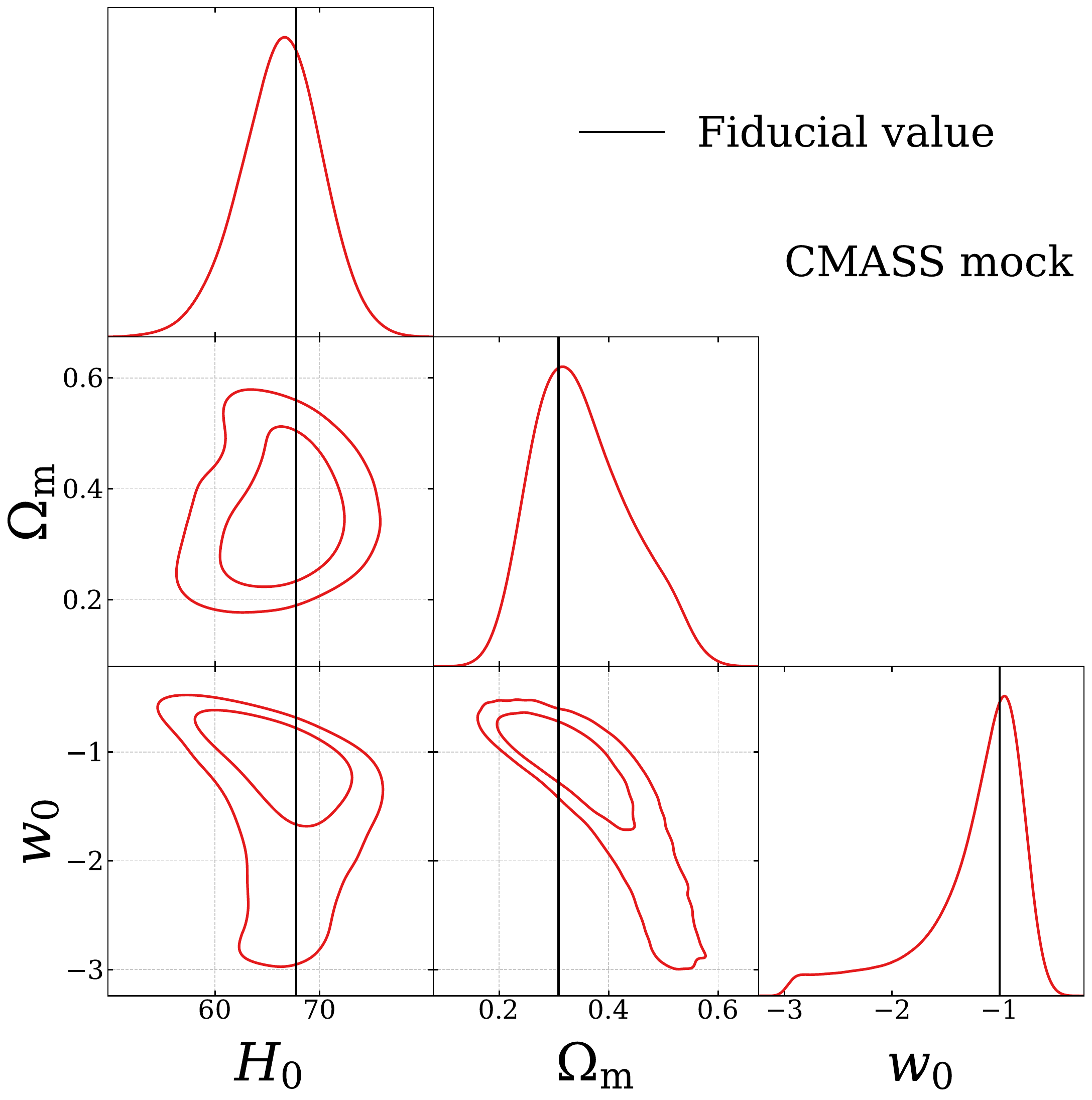}
\caption{\label{fig:CMASS_mock_H0Omw0} Similar to the right panel of Figure~\ref{fig:CMASS_mock}, but the free parameters are replaced by $H_0$, $\Omega_{\rm m}$ and $w_0$. The theoretical values are derived from the fiducial cosmology of the CMASS mock catalogs.}
\end{figure*}

To validate the ability of our model to accurately recover cosmological parameters from the observed galaxy power spectrum, we perform a pipeline test using the MultiDark-Patchy mock catalogs. The left panel of Figure~\ref{fig:CMASS_mock} displays the average and variance of the  measured power spectrum multipoles from the CMASS mocks, while the right panel shows the posterior distributions of the three cosmological observables -- $f$, $\alpha_\parallel$ and $\alpha_\perp$. All fiducial values lie within the 1$\sigma$ confidence regions, supporting the reliability of our model. {For the fittings of cosmological parameters ($H_0$, $\Omega_{\rm m}$ and $w_0$) shown in Figure~\ref{fig:CMASS_mock_H0Omw0}, we reach a consensus.}

\subsection{Tests on WebSky-CMASS}
\label{subapp:webskycmass_test}

\begin{figure*}[t!]
\includegraphics[width=0.48\textwidth]{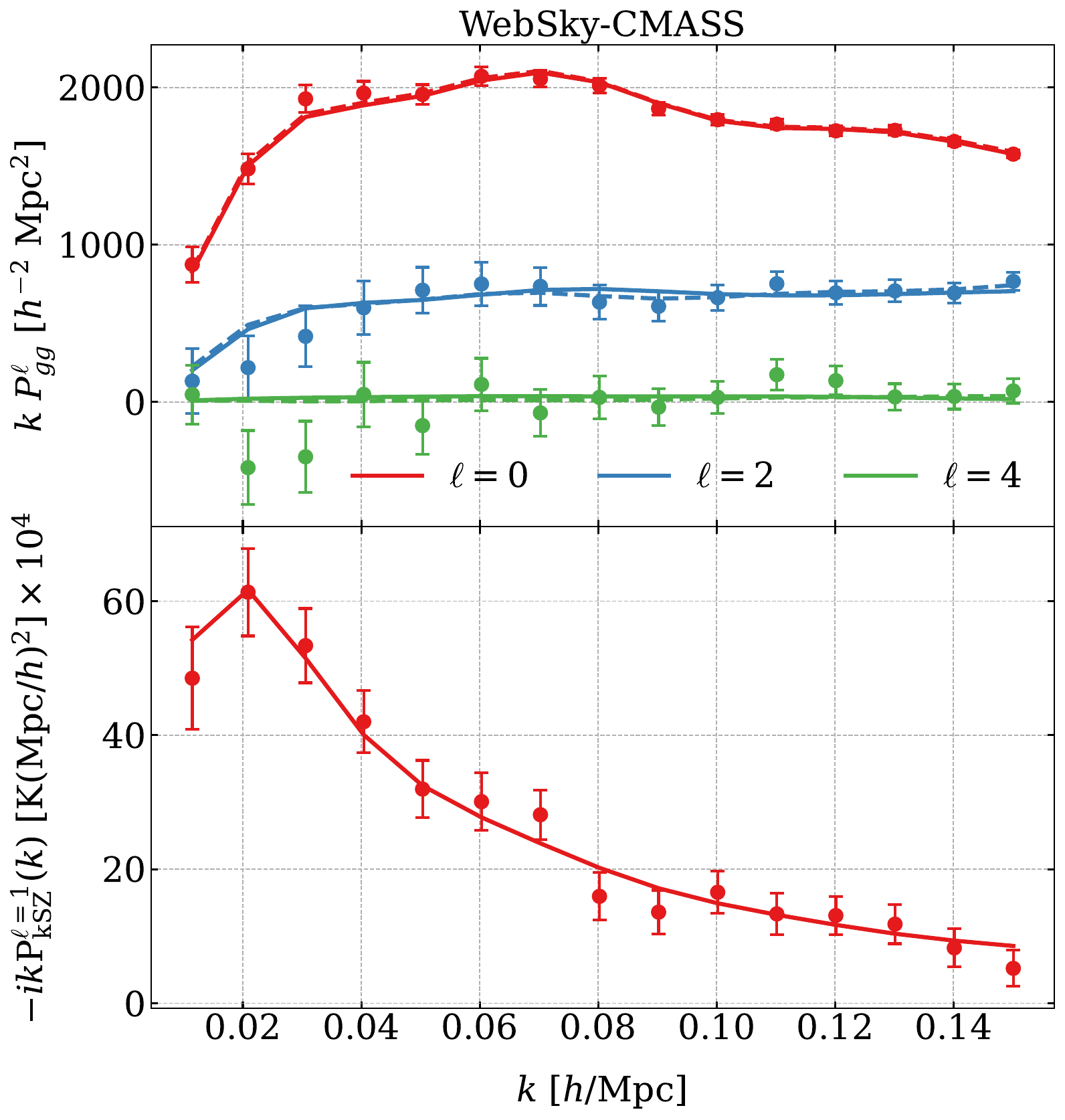}
\includegraphics[width=0.5\textwidth]{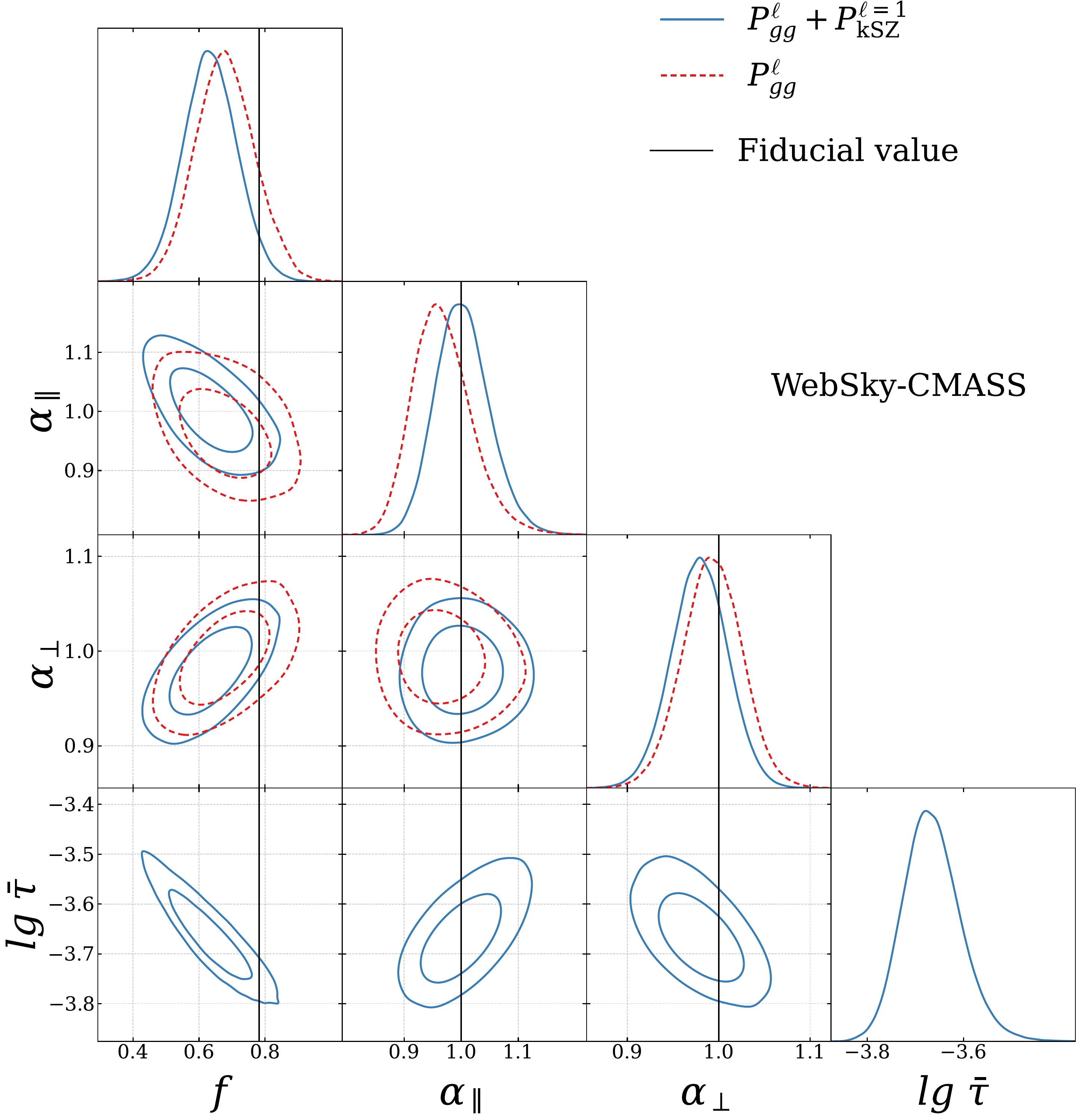}
\caption{\label{fig:websky_CMASS} The same as Figure~\ref{fig:CMASS}, but for WebSky-CMASS.}
\end{figure*}

\begin{figure*}[t!]
\centering
\includegraphics[width=0.45\textwidth]{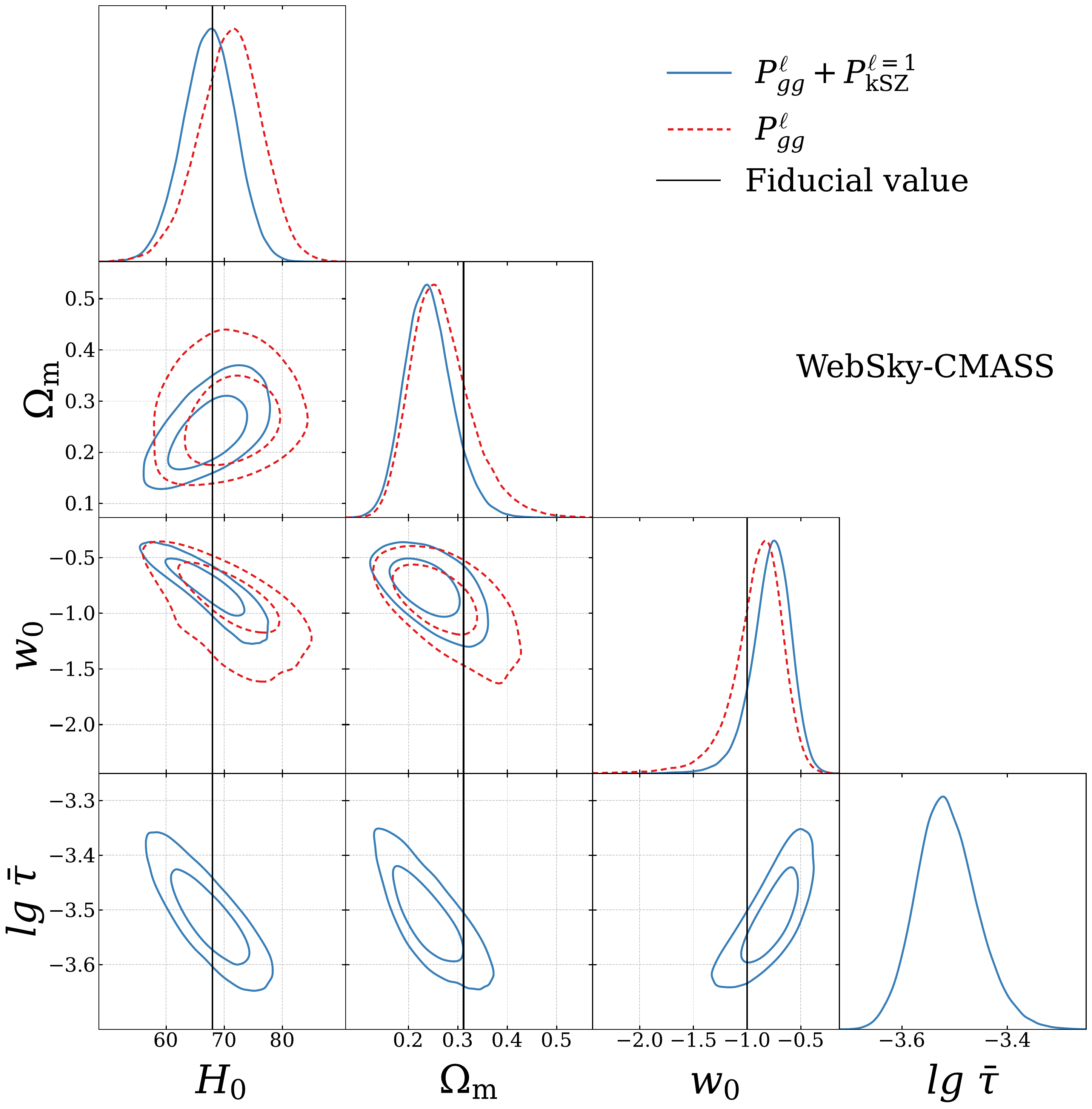}
\includegraphics[width=0.45\textwidth]{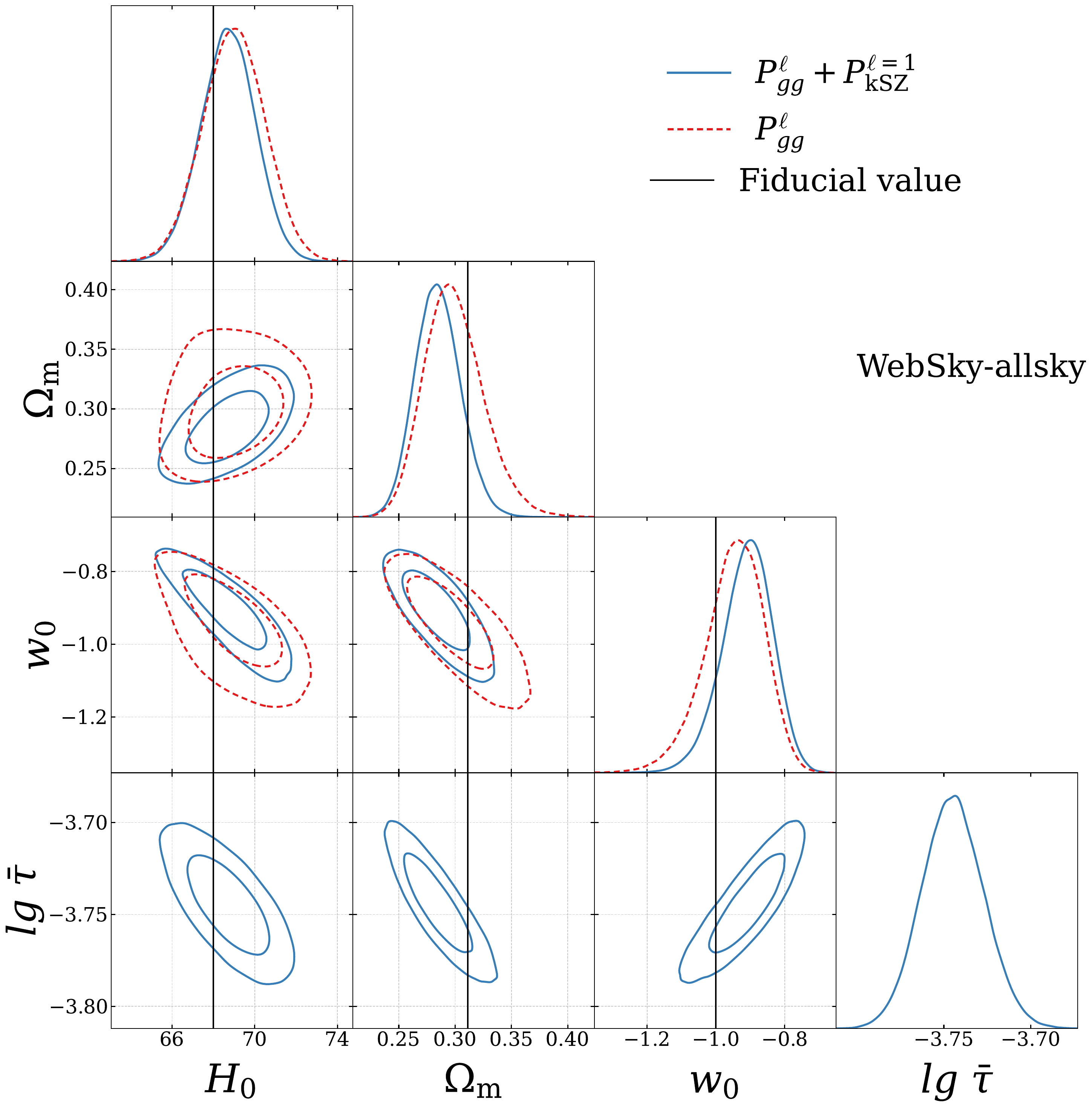}
\caption{\label{fig:websky_CMASS_H0Omw0} The same as Figure~\ref{fig:CMASS_H0Omw0}, but for WebSky-CMASS (\textit{left}) and WebSky-allsky (\textit{right}).}
\end{figure*}

{We employ the WebSky simulation to validate the theoretical model and to evaluate whether the joint analysis provides stronger constraints on cosmological parameters compared to using the galaxy power spectrum alone. Figure~\ref{fig:websky_CMASS} presents the results from the WebSky-CMASS analysis. The right panel displays the corresponding posterior distributions. The fitting results show that the fiducial values fall within or near the range of the posterior distributions by $1\sigma$. The FoM improvements for the parameter pairs $f-\alpha_\parallel$, $f-\alpha_\perp$ and $\alpha_\parallel-\alpha_\perp$ are \LSH{24.5\%, 19.1\%, and 11.4\%}, respectively. This clearly demonstrates that the kSZ power spectrum provides valuable additional information for cosmological parameter constraints.

The results of the constraints on the cosmological parameters are shown in the left of Figure~\ref{fig:websky_CMASS_H0Omw0}. The FoMs improve by \LSH{42.5\% for $H_0-\Omega_{\rm m}$, 50.6\% for $H_0-w_0$ and 29.0\% for $\Omega_{\rm m}-w_0$}. The improvement in the constraint ability of the cosmological parameters for the joint analysis is more significant compared to the cosmological observables.}

\subsection{Tests on WebSky-allsky}
\label{subapp:webskyallsky_test}
\begin{figure*}
\includegraphics[width=0.48\textwidth]{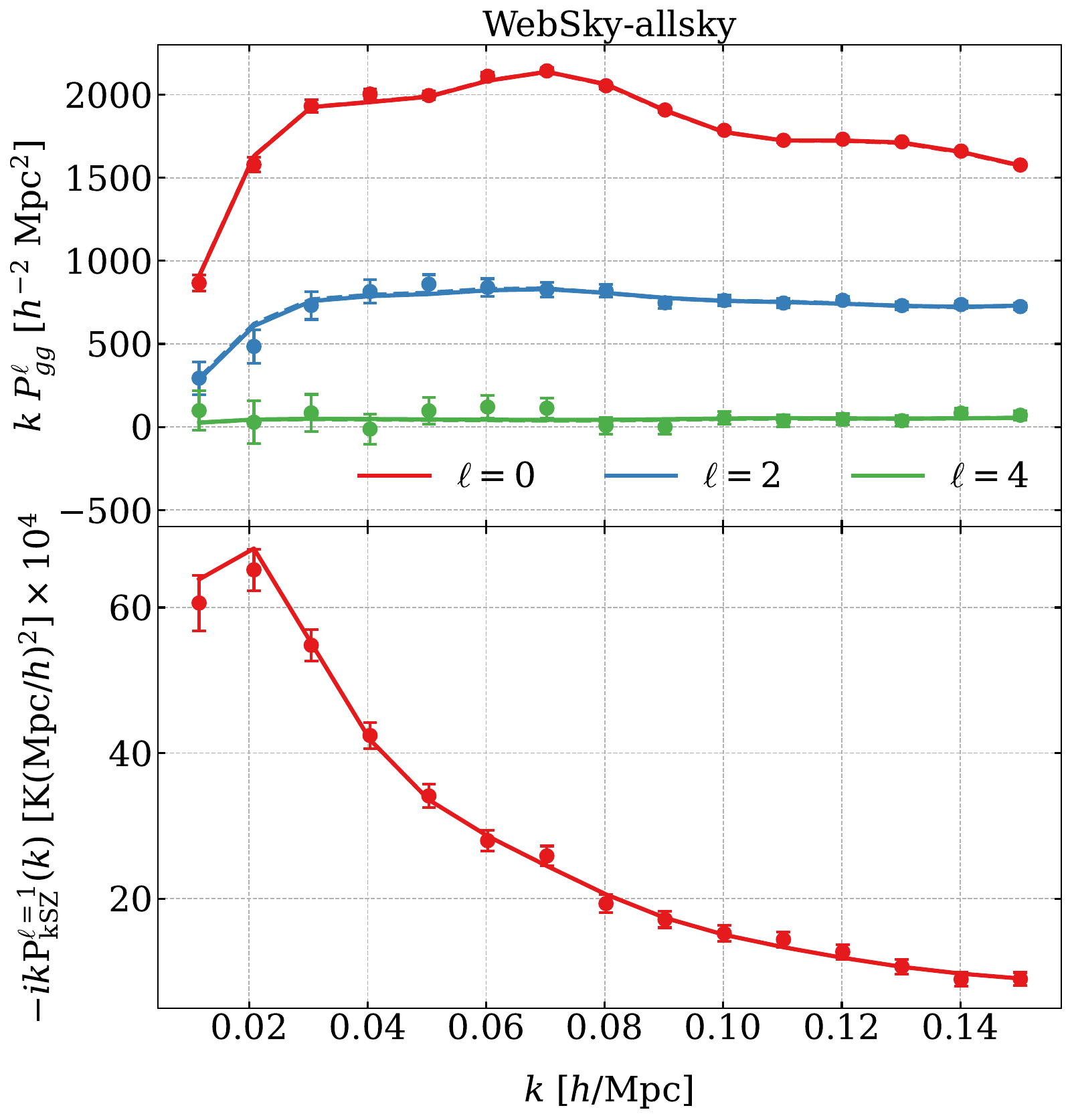}
\includegraphics[width=0.5\textwidth]{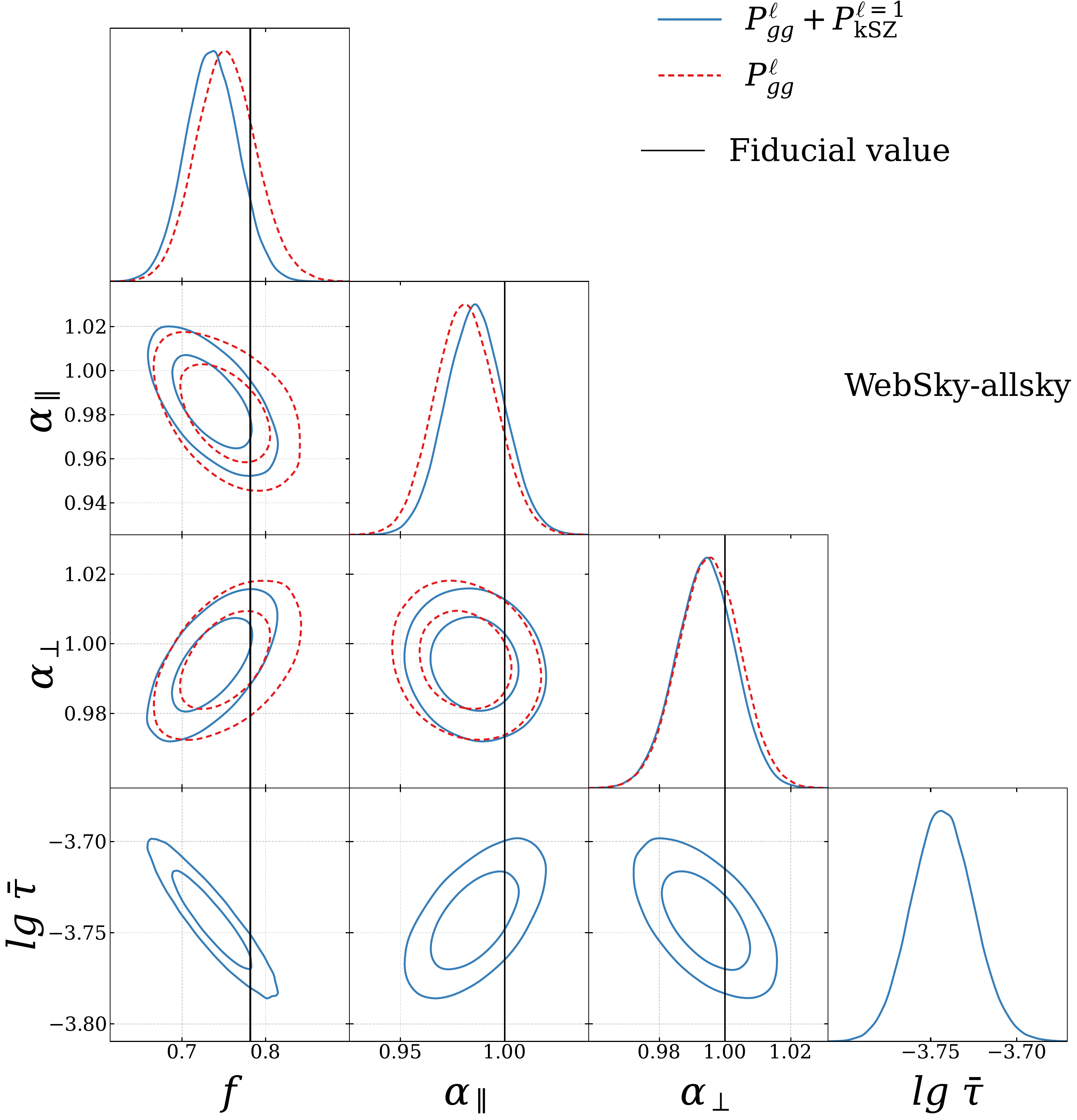}
\caption{\label{fig:websky_all_sky}  The same as Figure~\ref{fig:CMASS}, but for WebSky-allsky.}
\end{figure*}

{We further suppress the impact of cosmic variance by using the WebSky-allsky mock, which has a larger survey volume. It is useful for understanding the systematic errors induced by the inaccuracy of our adopted power spectrum models. The results from the full-sky WebSky-allsky sample are presented in Figure~\ref{fig:websky_all_sky}.  The FoMs improve by \LSH{24.3\% for $f-\alpha_\parallel$, 23.8\% for $f-\alpha_\perp$, and 8.7\% for $\alpha_\parallel-\alpha_\perp$}, demonstrating enhanced constraining power from the joint analysis. For the constraints of the cosmological parameters, the results are shown in the right of Figure~\ref{fig:websky_CMASS_H0Omw0}. The FoMs improve by \LSH{40.6\% for $H_0-\Omega_{\rm m}$, 43.3\% for $H_0-w_0$ and 31.2\%} for $\Omega_{\rm m}-w_0$. These improvements confirm that the joint analysis provides tighter constraints than using the galaxy power spectrum alone. In addition, the systematic differences between the fiducial and best-fitted values of WebSky-allsky are within $1\sigma$ error of WebSky-CMASS, showing that the theory model in this Letter is applicable to the CMASS+ACT data analysis.}

\section{Robustness Tests for the Covariance Matrix}
\label{app:cov_validation}
\begin{figure}[t!]
\centering
\includegraphics[width=0.35\linewidth]{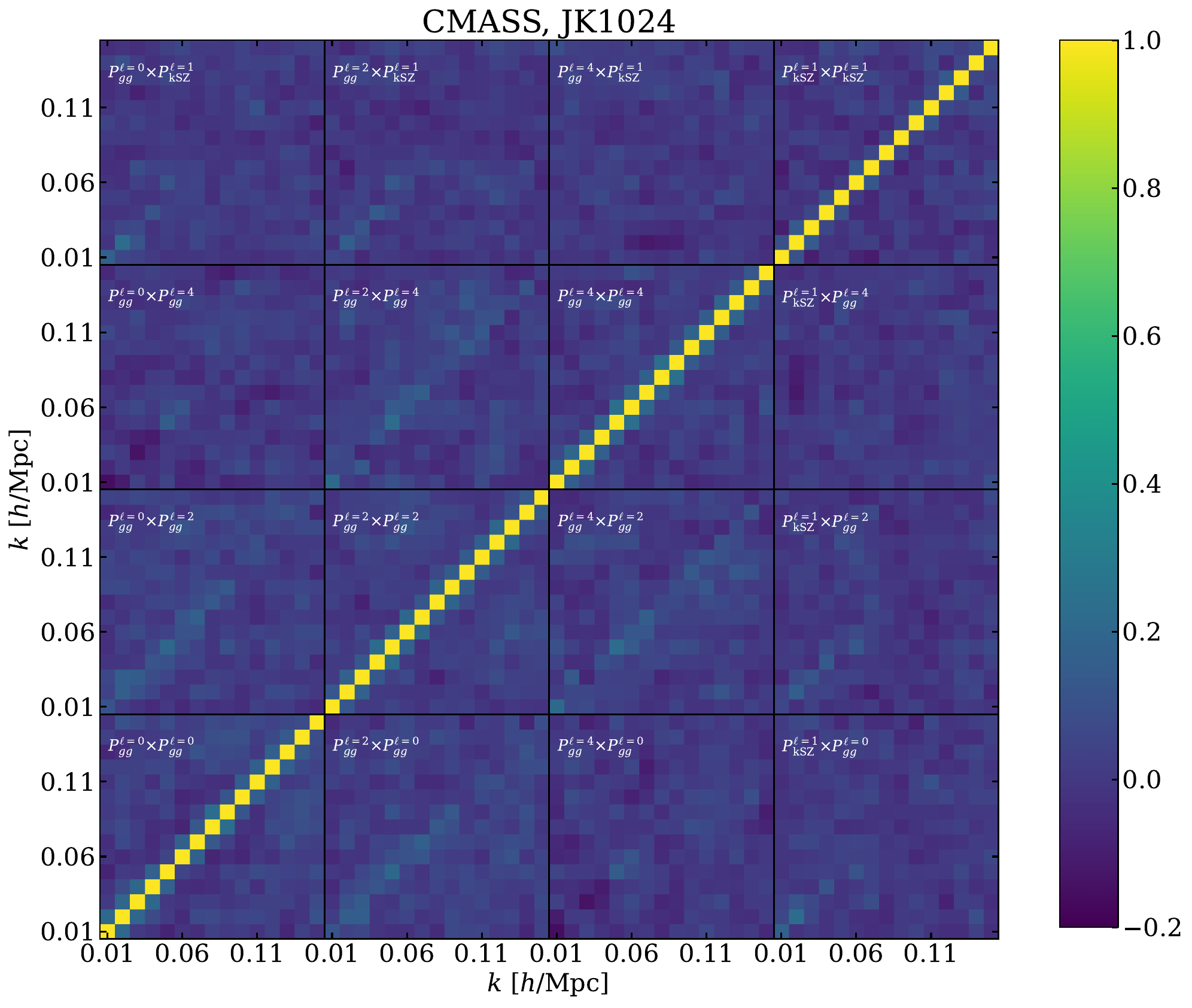}
\includegraphics[width=0.3\linewidth]{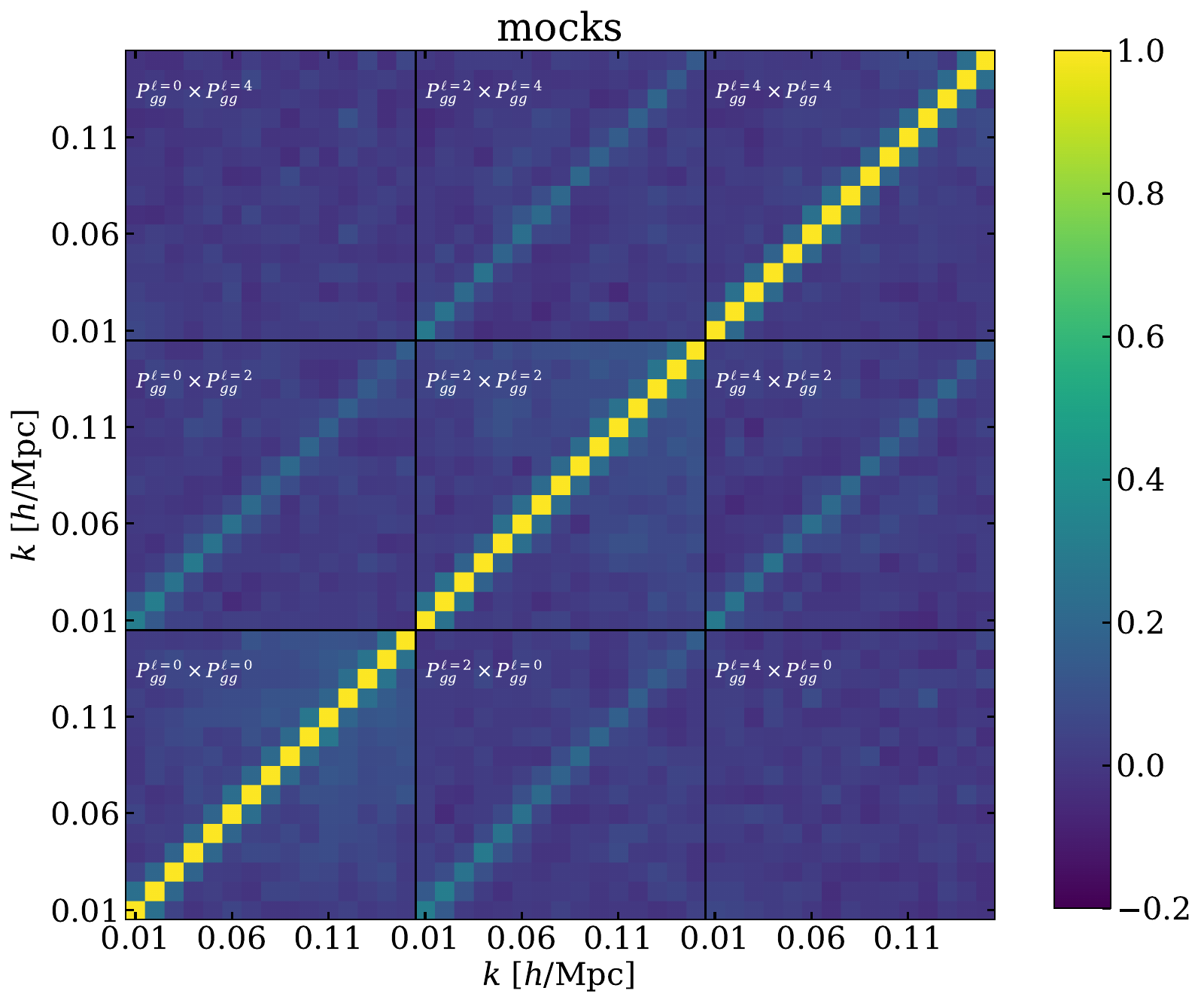}
\includegraphics[width=0.3\linewidth]{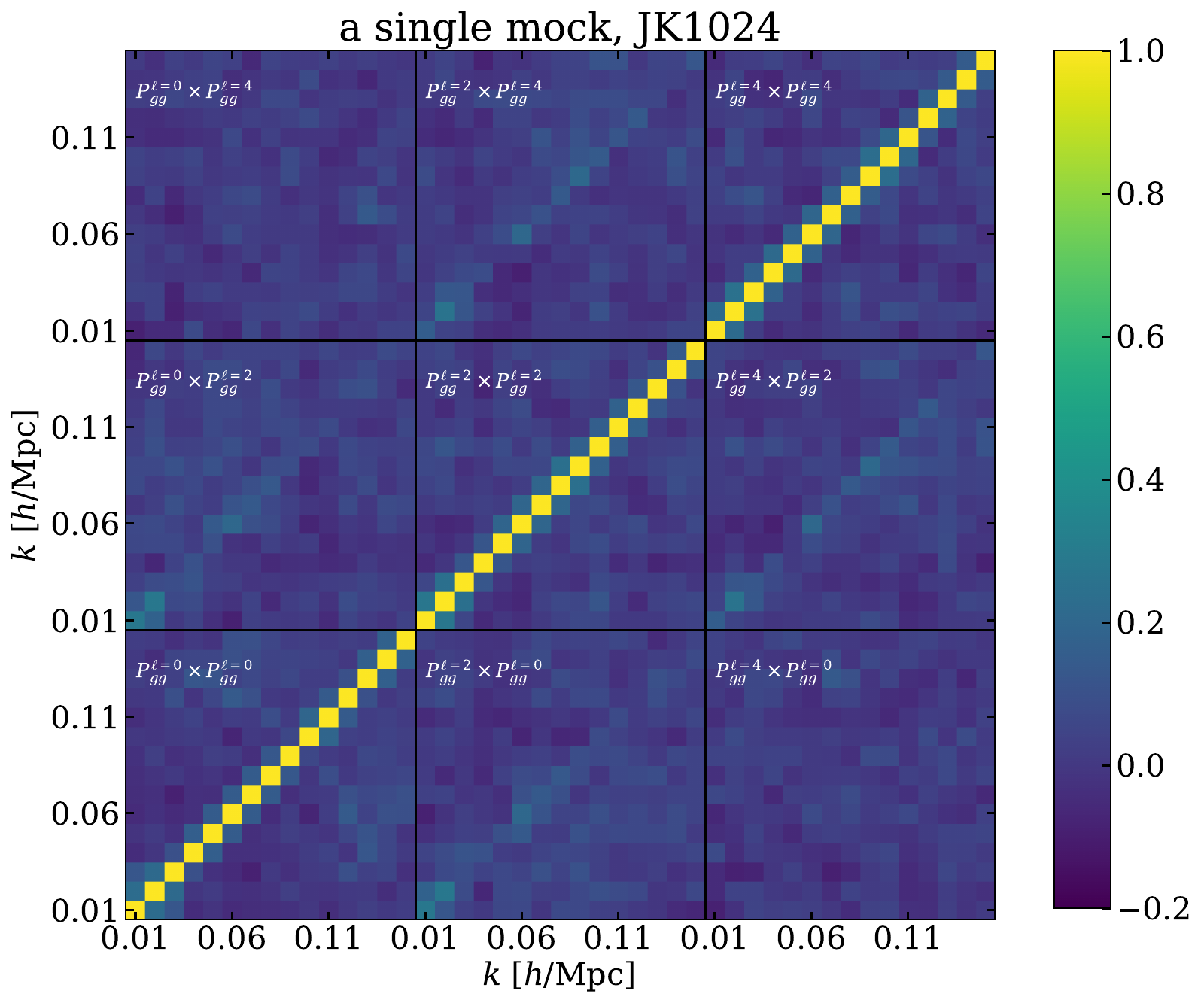}
\caption{\label{fig:corr_coeff} Correlation coefficient matrices derived from $\njk=1204$ JK resampling on the CMASS data (\textit{left}), the full set of 2048 CMASS mocks (\textit{middle}), and the $\njk=1204$ JK on a single CMASS mock (\textit{right}).}
\end{figure}
\begin{figure}[t!]
\centering
\includegraphics[width=0.48\linewidth]{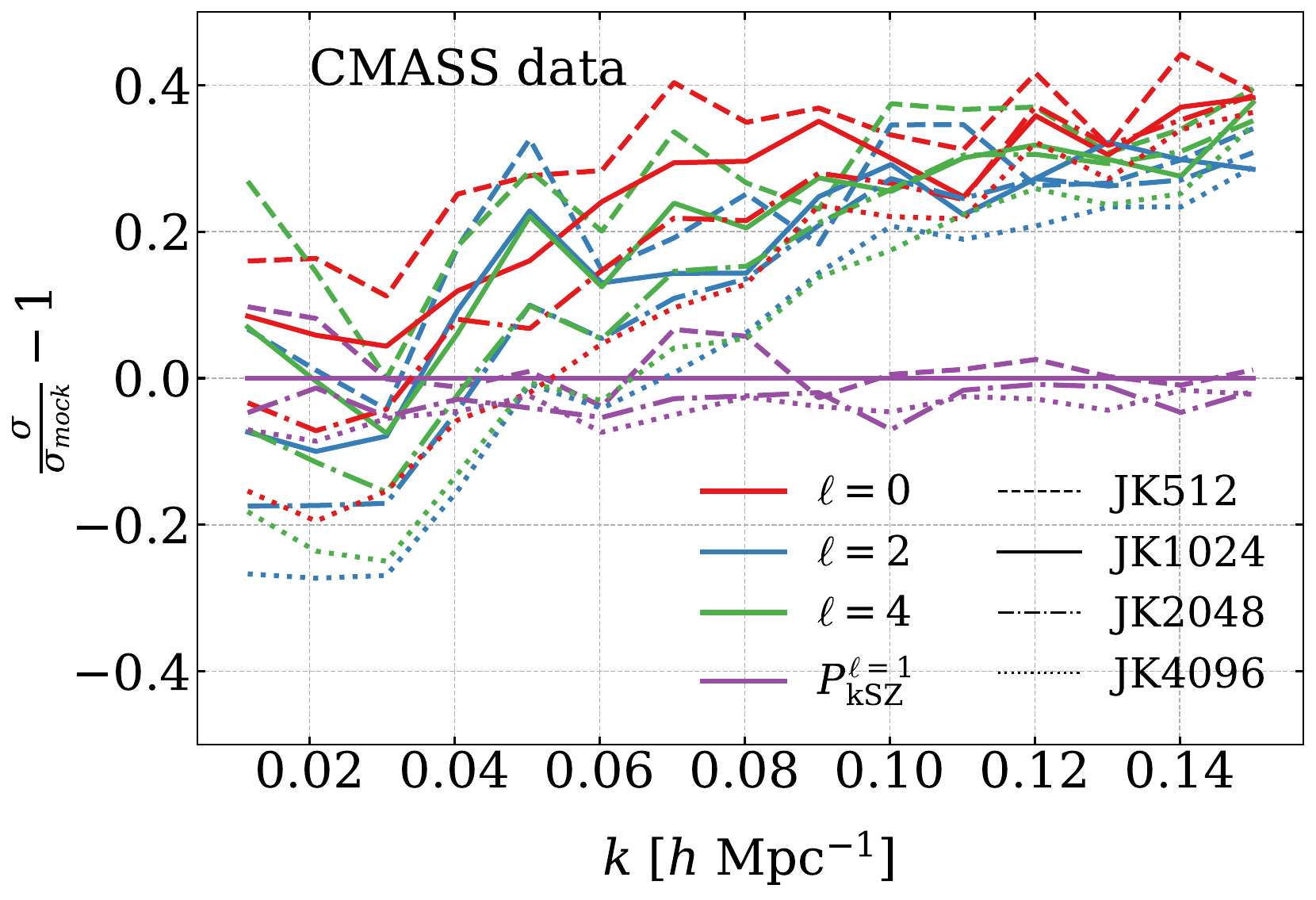}
\includegraphics[width=0.48\linewidth]{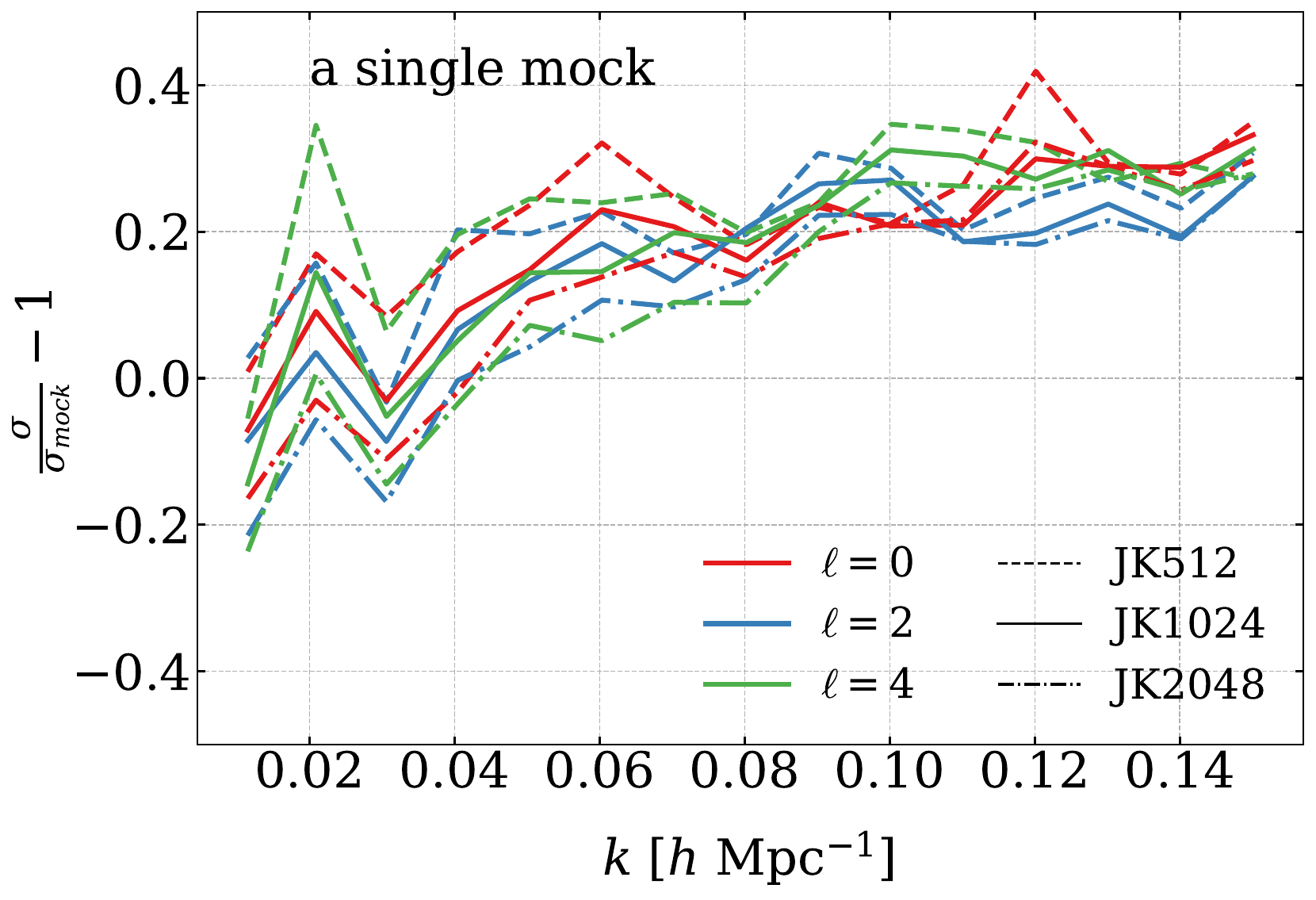}
\caption{\label{fig:cov_ratio} \textit{\bf Red, blue, green lines:} fractional differences between the JK-derived $P^{\ell}_{gg}$ variances and the mock-based ones. \textit{Left} is for the CMASS data and \textit{right} is for a randomly selected mock catalog. \textit{\bf Purple lines:} fractional differences between the JK-derived $P^{\ell}_{\rm kSZ}$ variances and the $N_{\rm JK}=1024$ JK-derived one.}
\end{figure}
\begin{figure}[t!]
\centering
\includegraphics[width=0.45\linewidth]{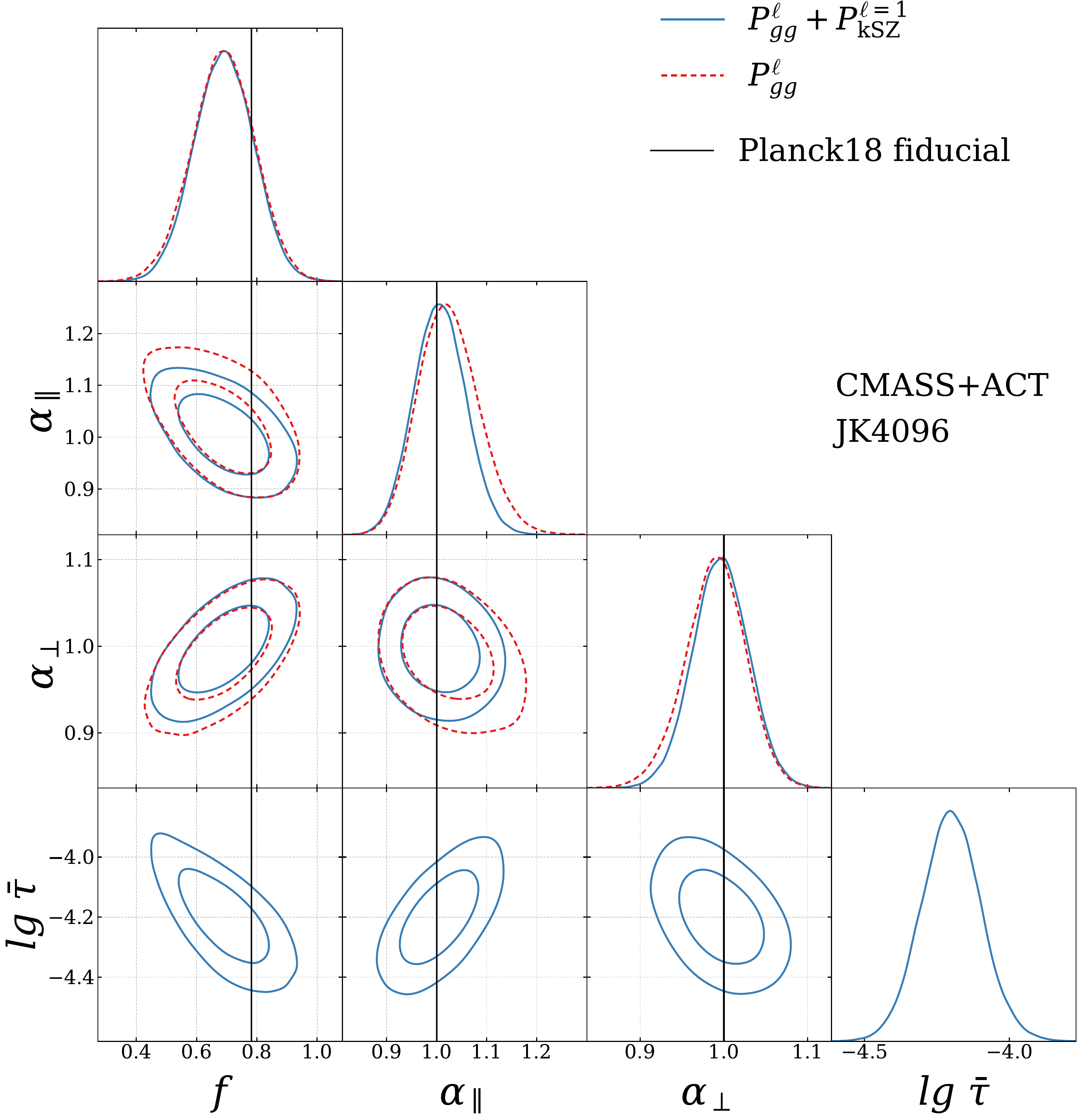}
\caption{\label{fig:mcmc_cmass_JKs} MCMC constraints on cosmological observables of CMASS+ACT data using $\njk=4096$ JK-derived covariance. The FoM improves by \LSH{$17.0\%$ for the $f$--$\alpha_\parallel$ pair, $11.1\%$ for $f$--$\alpha_\perp$, and $15.9\%$ for $\alpha_\parallel$--$\alpha_\perp$.}}
\end{figure}
\begin{figure}[t!]
\centering
\includegraphics[width=0.45\linewidth]{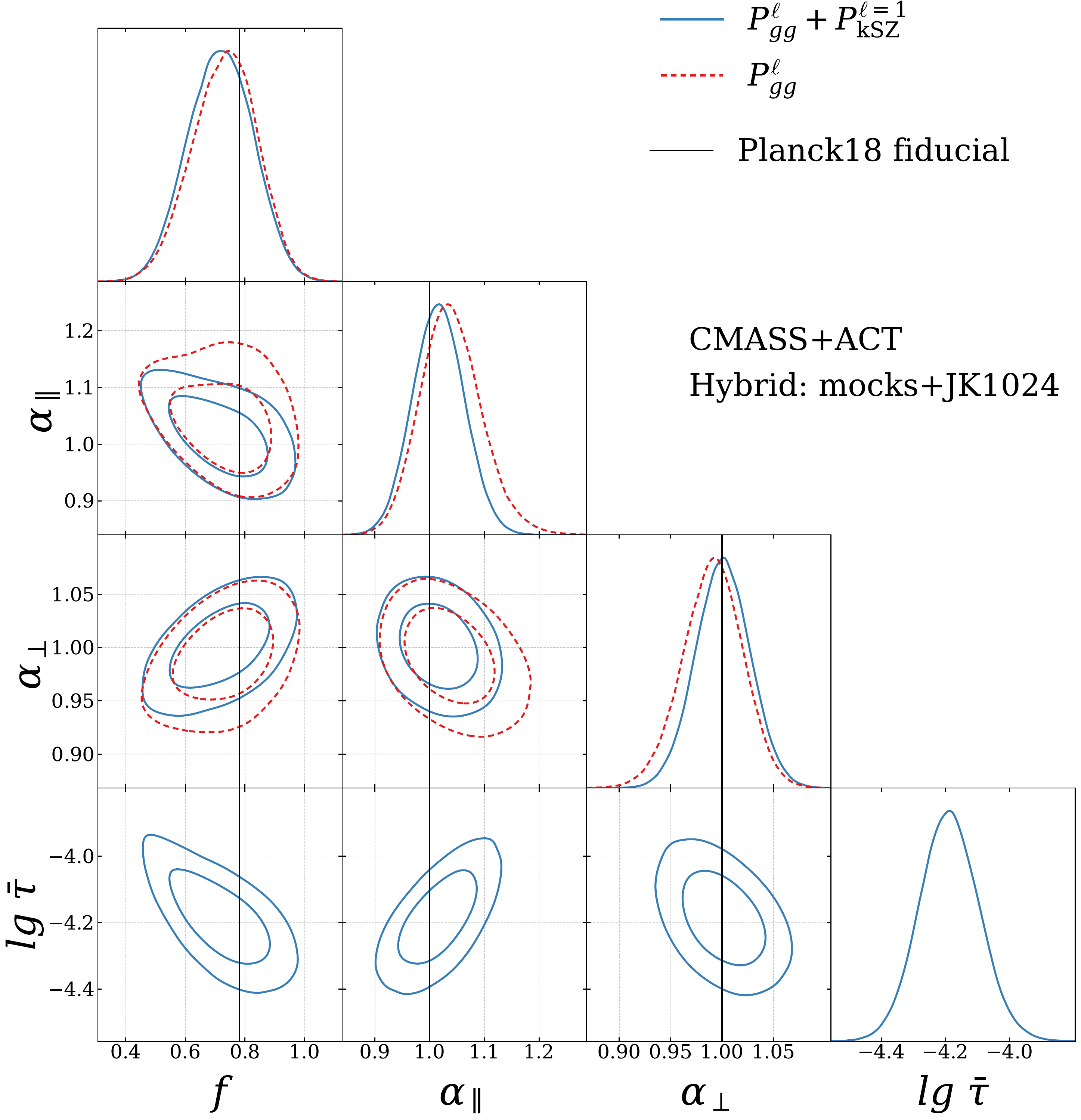}
\includegraphics[width=0.45\linewidth]{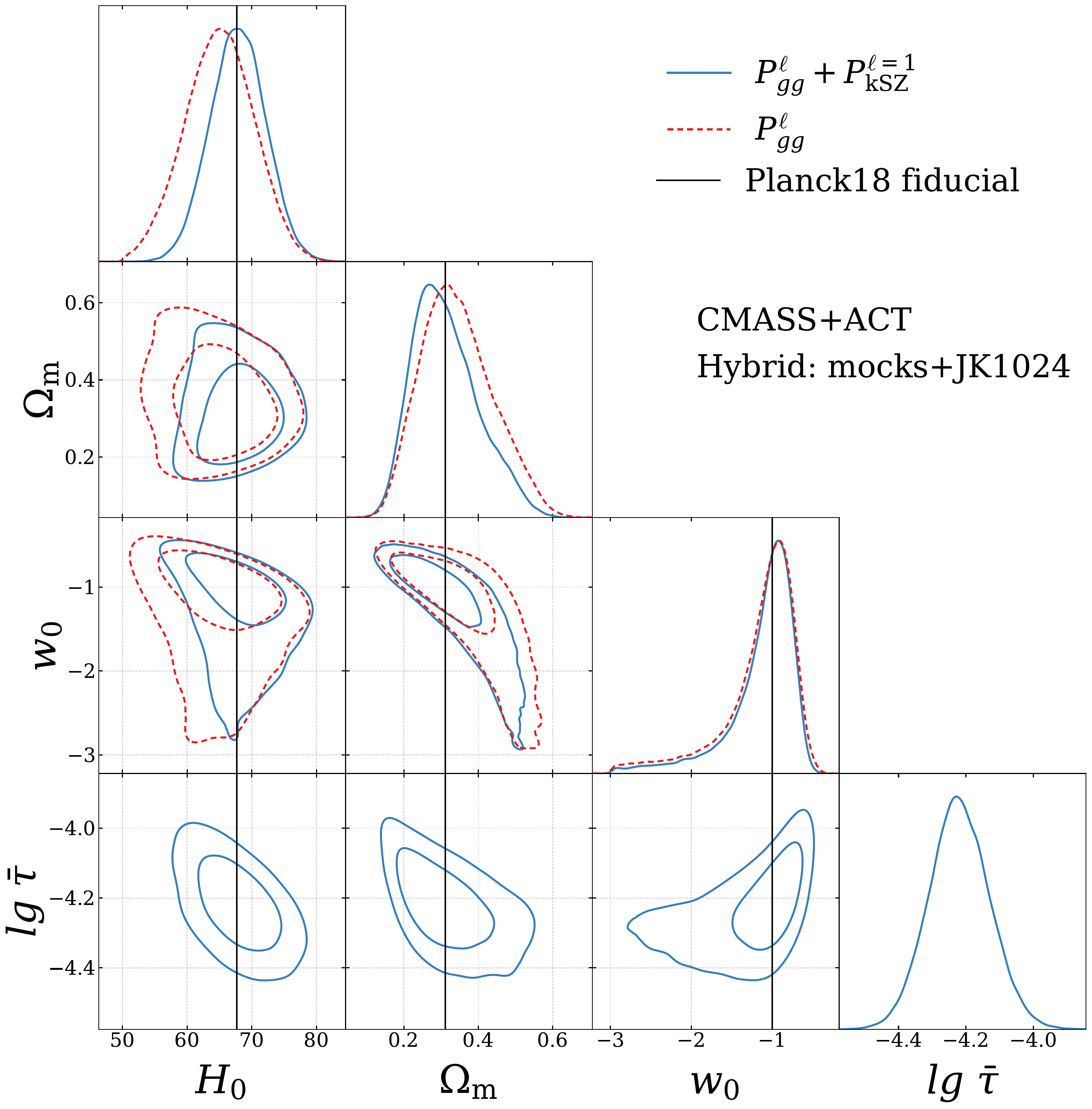}
\caption{\label{fig:corner_CMASS_Cov_hybrid} Results of the CMASS+ACT MCMC analysis using the hybrid covariance.  \textit{Left:} MCMC constraints on cosmological observables on CMASS+ACT data. The FoM improves by \LSH{$23.0\%$ for the $f$--$\alpha_\parallel$ pair, $15.4\%$ for $f$--$\alpha_\perp$, and $19.5\%$ for $\alpha_\parallel$--$\alpha_\perp$. \textit{Right} MCMC constraints on cosmological parameters of CMASS+ACT data. The FoM improves by $29.2\%$ for the $H_0$--$\Omega_{\rm m}$ pair, $34.7\%$ for $H_0$--$w_0$, and $28.4\%$ for $\Omega_{\rm m}$--$w_0$.}}
\end{figure}

\begin{figure}[t!]
\centering
\includegraphics[width=1\linewidth]{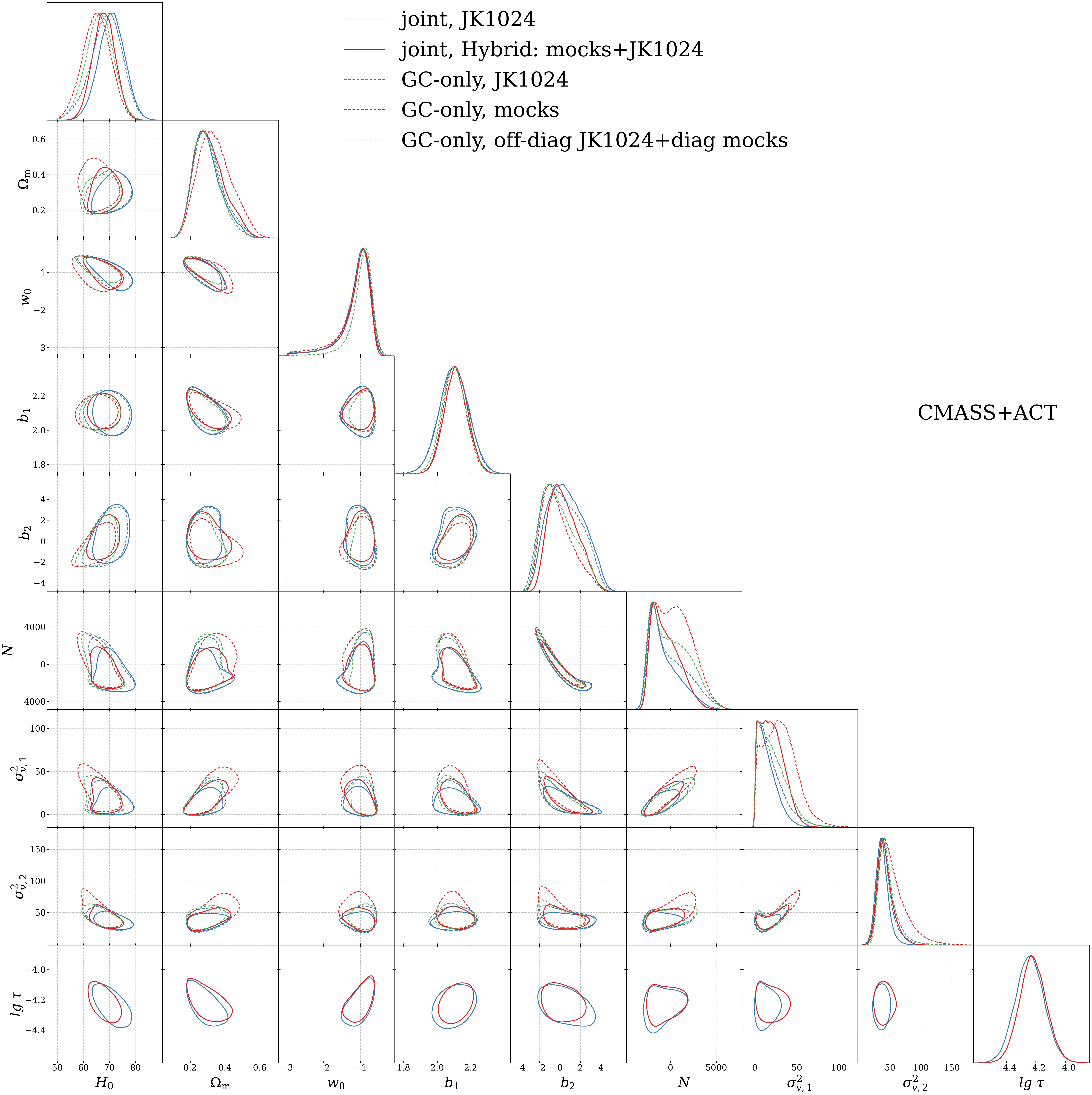}
\caption{\label{fig:corner_1sigma_cmass_paras_gconly+joint} Results of the CMASS+ACT MCMC analysis using different covariances. Only $1-\sigma$ contours are shown.  When adopting the hybrid covariance matrix, we apply Hartlap factor 0.940 for the JK1024 portion of the precision matrix $(1024-60-2)/(1024-1)$ and 0.970 for the mock portion $(2048-60-2)/(2048-1)$. Although this treatment is not very rigorous, considering the two factors (0.94 and 0.97) are close to each other, we do not expect it to affect the main conclusion of this work.}
\end{figure}
\begin{figure}[t!]
\centering
\includegraphics[width=1\linewidth]{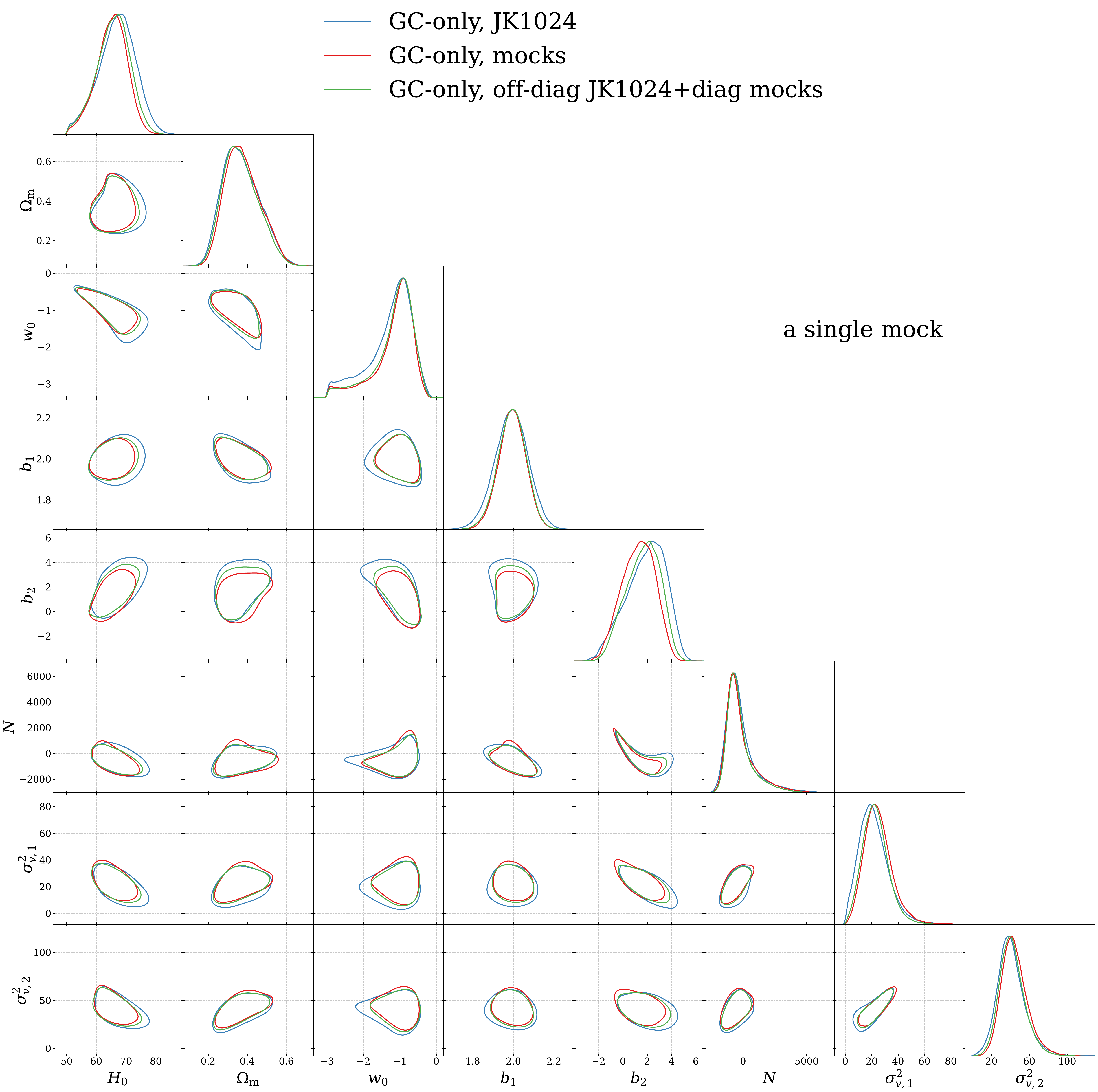}
\caption{\label{fig:corner_1sigma_mock_paras_gconly} MCMC results of the GC-only analysis on a single Patchy mock with different covariances. Only $1-\sigma$ contours are shown. When adopting the combined covariance matrix (`off-diag JK1024+diag mocks'), we apply a Hartlap factor of 0.978 for the diagonal elements of the precision matrix $(2048-45-2)/(2048-1)$ and 0.955 for the off-diagonal portion $(1024-45-2)/(1024-1)$. Although this treatment is not very rigorous, considering the two factors (0.978 and 0.955) are close to each other, we do not expect it to affect the main conclusion of this work.}
\end{figure}

{In this work, the covariance matrix for the CMASS data analysis is constructed via a data-driven approach: the delete-one JK method with $\njk=1024$. The robustness of this choice will be validated in this appendix, in comparison with the mock-based covariance matrix estimated from 2048 MultiDark-Patchy mocks.

In Figure~\ref{fig:corr_coeff}, we first show 2D plots of the correlation coefficient matrices derived from JK resampling on the CMASS data, the full set of 2048 CMASS mocks, and the JK on a single CMASS mock. As can be seen, the mock-based covariance (\textit{middle}) has larger off-diagonal elements than those of the JK-derived covariance (\textit{left/right}), and the latter elements seem noisier than the former ones.

\subsection{Tests on the $\njk=1024$ choice}
\label{subapp:njk_num}

In Figure~\ref{fig:cov_ratio}, we compare the multipole variances of the power spectrum derived from the JK method with those from the mock-based covariance. To show this, we compute the fractional differences of multipole variances for both the CMASS data (\textit{left}) and a randomly selected Patchy mock catalog (\textit{right}).

For the $P_{gg}$ multipoles, the JK method overestimates the power spectrum errors on most scales except the largest ones. This likely stems from the lack of strict independence among the $N_{\rm JK}$ samples -- a behavior consistent with earlier studies~\citep{Norberg2009,Favole2021}. As $N_{\rm JK}$ increases, the size of each subsample shrinks, progressively reducing the ability of the JK samples to capture the cosmic variance contribution to the multipole variance. Consequently, the JK-derived variance decreases with larger $N_{\rm JK}$ without showing clear convergence, especially on large scales.

When adopting the fiducial $N_{\rm JK}=1024$, the JK-derived variances at large scales agree most closely with the mock-based estimates. This agreement is one of the key reasons why we choose $N_{\rm JK}=1024$ as the default subsample number in our JK implementation.

A similar comparison for $P_{\mathrm{kSZ}}^{\ell=1}$ is shown by the purple curves in the left panel of Figure~\ref{fig:cov_ratio}. The $P_{\mathrm{kSZ}}^{\ell=1}$ variance shows better convergence across different $N_{\mathrm{JK}}$ choices, but it still decreases at large scales when $\njk$ increases, up to $10\%$, for the same reason presented before. This better convergence can be understood by noting that the sample covariance of the kSZ signal is dominated by residual primary CMB fluctuations and detector noise. For instance, the typical kSZ signal amplitude is $\mathcal{O}(0.1)\,\mu\mathrm{K}$, whereas the residual CMB and detector noise reach $\mathcal{O}(1\text{--}10)\,\mu\mathrm{K}$. As a result, even small JK subsamples remain effectively independent, and the estimated variance decreases slower with increasing $N_{\mathrm{JK}}$.

In turn, because both the $P^\ell_{gg}$ and $P^{\ell=1}_{\rm kSZ}$ variances decrease as $\njk$ increases, we must examine how the improvement in the FoM depends on the choice of $\njk$. Figure~\ref{fig:mcmc_cmass_JKs} presents the MCMC results with $\njk=4096$. Comparing with Figure~\ref{fig:CMASS} in which $\njk=1024$, the overall FoM improvement remains stable across the two $\njk$ values, confirming that our main conclusion is robust to the choice of the JK subsample number. 

\subsection{Comparison with the mock-based covariance}
\label{subapp:cov_comparison}

Next, as noted in Figure~\ref{fig:cov_ratio}, the JK-derived $P_{gg}^{\ell}$ covariance tends to overestimate the uncertainty in $P_{gg}^{\ell}$. To address this, we test a hybrid covariance scheme: adopting the mock-based covariance for $P_{gg}$ while keeping the JK covariance for $P_{\rm kSZ}$ and the cross-terms. The results are shown in Figure~\ref{fig:corner_CMASS_Cov_hybrid}. Intriguingly, this hybrid approach leads to a larger improvement in the FoM than our original baseline. To understand this, we compare the $1\sigma$ contours obtained with the JK-derived and hybrid covariances in Figure~\ref{fig:corner_1sigma_cmass_paras_gconly+joint}. For brevity, here we show only the cosmological parameter contours, as the results for cosmological observables are similar. In the same figure, we also include a third contour (the green dashed line) computed from a combined covariance that uses the diagonal elements of the mock-based covariance and the off-diagonal elements of the JK-based correlation coefficient matrix. This covariance helps us study the impact of the covariance off-diagonal elements on the fitting results.

By comparing the red and blue contours in Figure~\ref{fig:corner_1sigma_cmass_paras_gconly+joint}, we observe that the joint analysis contours derived from the hybrid covariance are smaller than those from the JK-derived covariance. In contrast, for the GC-only analysis, the hybrid contours are comparable to or even larger than their JK counterparts, and their degeneracy directions also shift, which is an unexpected outcome. Along with this, when using the mock-based covariance (red dashed contours), although the GC-only constraints on the nuisance parameters $b_1$ and $b_2$ tighten, those on $N$, $\sigma_{v,1}^2$, and $\sigma_{v,2}^2$ become significantly larger. Yet by including the kSZ power spectrum (red solid contours), these weird behaviors of the nuisance parameter constraints disappear and the constraints on cosmological parameters are back to expectations.

A further comparison between the red dashed and green dashed contours in Figure~\ref{fig:corner_1sigma_cmass_paras_gconly+joint} indicates that the off-diagonal elements of the two covariance matrices are primarily responsible for the observed behavior. To examine this further, we repeate the same test on a randomly selected mock catalog, as shown in Figure~\ref{fig:corner_1sigma_mock_paras_gconly}. Since the mock contains no kSZ signal, only the GC‑only analysis is displayed. Notably, no mismatch appears among the contours derived from the three covariance choices; instead, as expected, the contours gradually shrink when moving from the JK-derived to the combined and then to the mock-based covariance.

The contrast between Figures~\ref{fig:corner_1sigma_cmass_paras_gconly+joint} and~\ref{fig:corner_1sigma_mock_paras_gconly} reveals a subtle but important point: the anomaly in Figure~\ref{fig:corner_1sigma_cmass_paras_gconly+joint} is not simply due to differences between the JK‑derived and mock‑based covariances of the same dataset. Rather, it stems from a discrepancy between the off‑diagonal structure of the mock‑based covariance and the true covariance of the actual data. In other words, the mock catalog appears to fail in capturing certain details of the real observations. For example, there may be a mismatch between the Patchy mocks and the CMASS data in describing higher-order GC statistics, such as the trispectrum. This mismatch prevents the GC‑only analysis from robustly constraining the nonlinear nuisance parameters. When kSZ information is incorporated, however, those parameters become well constrained and the anomaly disappears.

To further exclude potential systematic influences, we compare MCMC results with the chain length being $8\times10^5$ steps and then tripled to $2.4\times10^6$ steps, while the latter number is adopted in all MCMC analysis in this Letter. The key features described above remain unchanged. We also repeated the GC‑only analysis using a cosmology consistent with that of the Patchy mocks, in order to rule out effects from differences in the fiducial cosmology adopted for the CMASS data. The results are qualitatively the same, confirming that a cosmology mismatch is not the cause of the observed anomaly. Furthermore, we repeat all tests in this appendix using $\njk=2048$ so that the Hartlap factor for the hybrid or combined covariance is simply the one for $2048$ samples. Again, we observe nearly the same results. For brevity, these results are not shown here.

Given these findings, and since the primary goal of this Letter is to evaluate improvements in cosmological constraints from including kSZ information, we judge it preferable to use a consistent covariance treatment for the GC, kSZ, and their cross-terms. We therefore adopt the  $\njk=1024$ JK-derived covariance for all components in the main analysis.}

\section{Mass distribution of WebSky-CMASS halos}
\label{app:mass_dis_websky}
\begin{figure}[t!]
\centering
\includegraphics[width=0.45\textwidth]{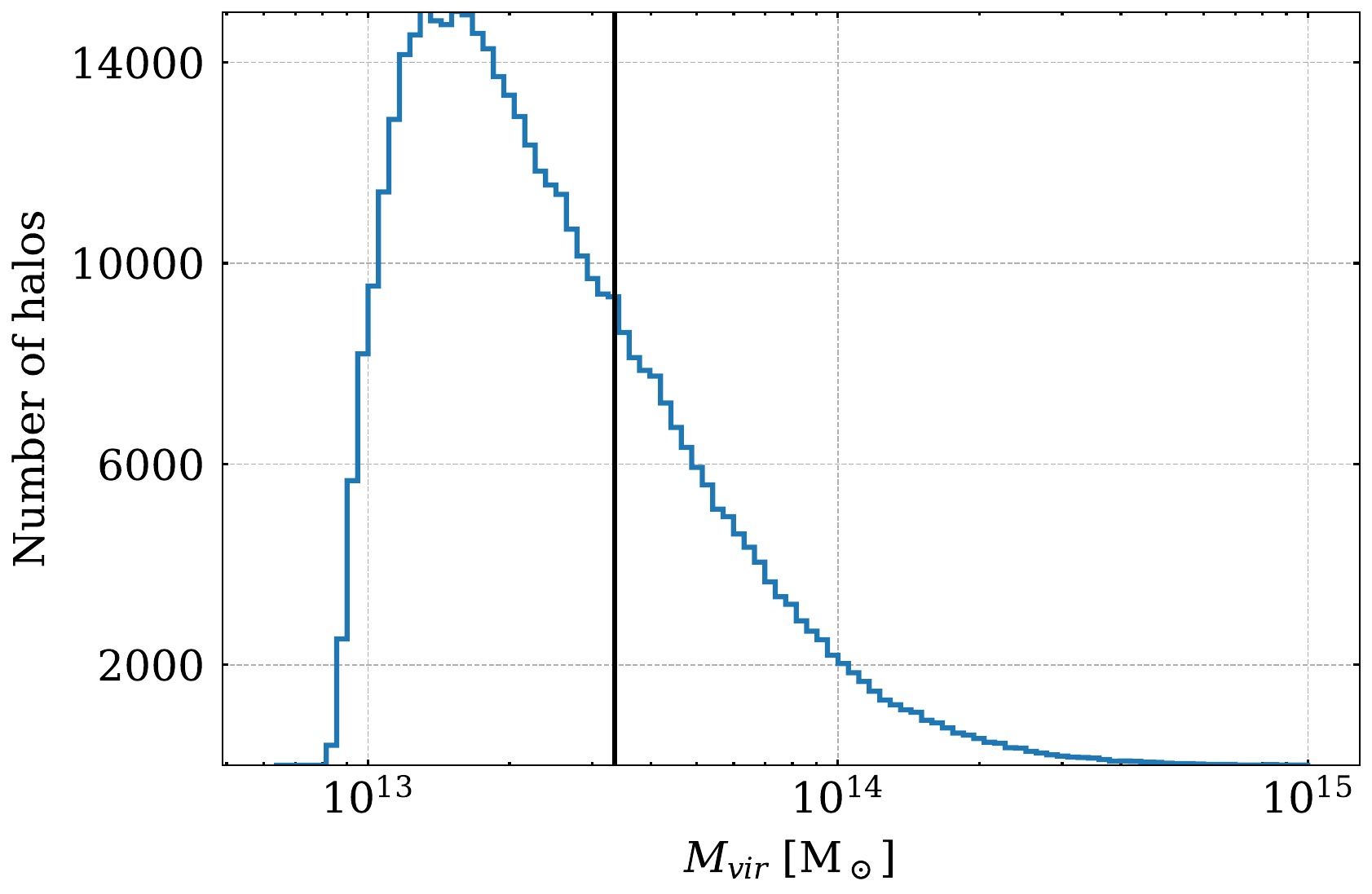}
\caption{\label{fig:mass_distribution_websky_CMASS} Mass distribution derived from WebSky-CMASS. The mean halo mass, indicated by the black vertical line, is {$3.4\times10^{13}\rm M_\odot$}, which corresponds to {$4.9\times10^{13}h^{-1}\rm M_\odot$ }when expressed in units of $h^{-1}\rm M_\odot$.}
\end{figure}

This appendix is relevant for explaining the constrained optical depth difference between the CMASS+ACT and Websky mocks. We convert the $M_{\rm 200m}$ values from the WebSky-CMASS simulation to $M_{\rm vir}$ under the assumption of an Navarro–Frenk–White (NFW) profile~\cite{NFW1997} for the dark matter halo distribution. The conversion incorporates the mean redshift of the sample and the concentration-mass relation from \cite{Duffy2008}, as implemented in the \texttt{Colossus} code package\footnote{\href{https://bdiemer.bitbucket.io/colossus/halo_concentration.html}{https://bdiemer.bitbucket.io/colossus/halo\_concentration.html}}. As shown in Figure~\ref{fig:mass_distribution_websky_CMASS}, the resulting halo mass distribution in WebSky-CMASS is systematically higher than that derived from the observed CMASS sample (Figure~3 of~\cite{Schaan2021}). {In particular, the mean halo mass in WebSky-CMASS is {$4.9 \times 10^{13} \, h^{-1} M_\odot$}, nearly twice the value of $2.6 \times 10^{13} \, h^{-1} M_\odot$ estimated for CMASS~\citep{White2011}.}

\section{Prediction of the Fisher Matrix}
\label{app:fisher_matrix}
\begin{figure*}
\includegraphics[width=1.0\textwidth]{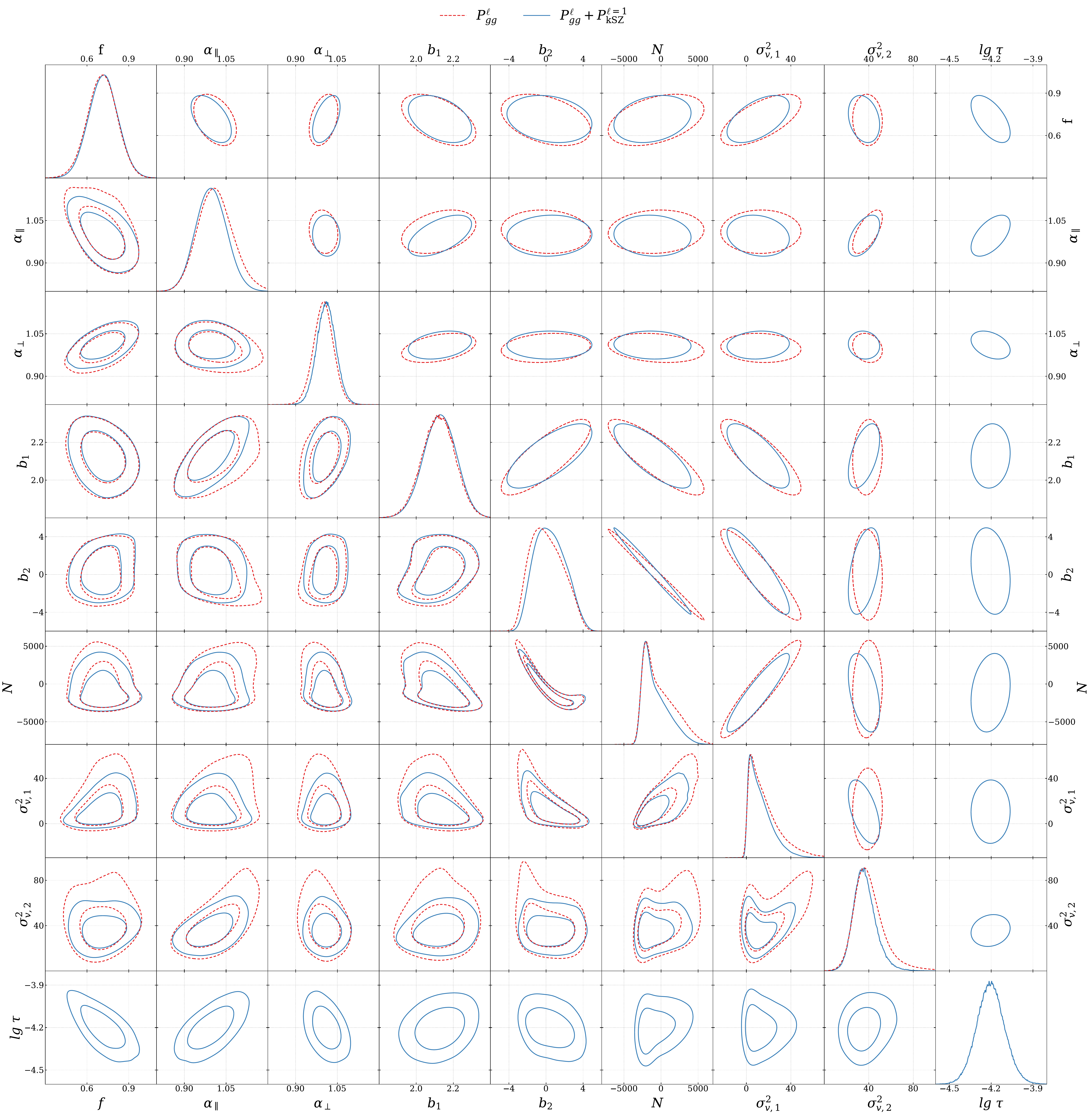}
\caption{\label{fig:corner_CMASS_all_parameters} \textit{Lower left panels:} the posterior distributions of all parameters, including the cosmological observables, constrained by the CMASS+ACT data. The diagonal panels show the corresponding one-dimensional marginalized distributions. \textit{Upper right panels:} the $1\sigma$ confidence levels predicted by the Fisher matrix formalism. The predictions in the upper right panels use corresponding fiducial values from the lower left panels.}
\end{figure*}

\begin{figure*}
\includegraphics[width=1.0\textwidth]{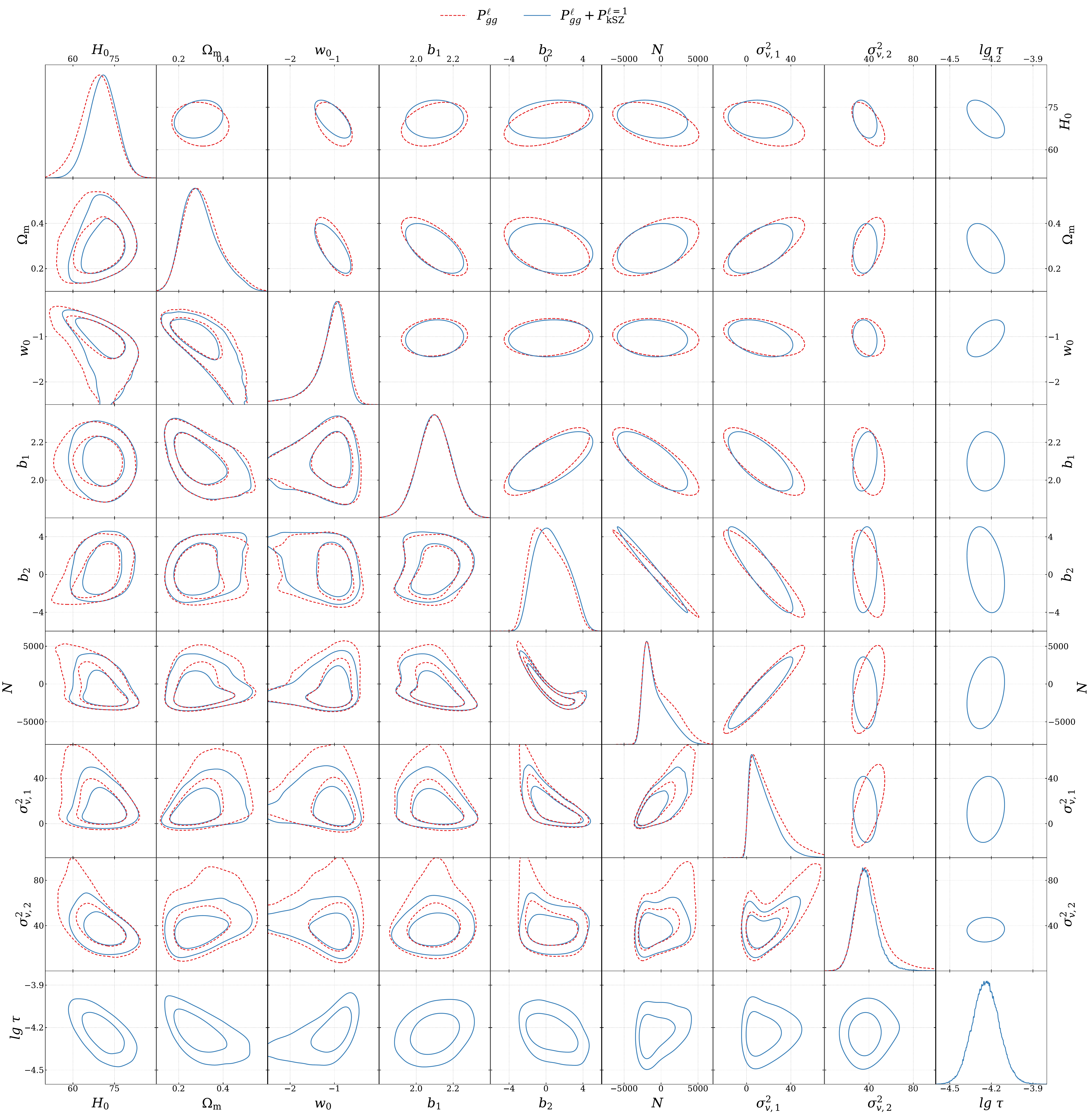}
\caption{\label{fig:corner_CMASS_all_parameters_H0Omw0} The same as Figure~\ref{fig:corner_CMASS_all_parameters}, but for the cosmological parameters. }
\end{figure*}

To quantify the constraining power of the GC-only and joint analyses and to investigate the origin of the differing improvements in the FoMs between cosmological observables and parameters (as shown in Section~\ref{sec:results}), we employ the Fisher matrix formalism. The Fisher matrix is given by
\begin{equation}
\label{eq:fisher_matrix}
F_{\alpha\beta}=\sum_{i,j=1}^{45(60)}\frac{\partial P_i}{\partial \theta_\alpha}[{\rm Cov}^{-1}]_{ij}\frac{\partial P_j}{\partial \theta\beta} + (prior),
\end{equation}
where ${\theta_\alpha}$ represents the set of parameters, with $\alpha=1,...,8 (9)$ for the GC-only (joint) analysis and $(prior)$ being the prior term. Here, $P_i$ corresponds to $P_{gg}^\ell$ (45 k-modes) or $P_{gg}^\ell+P_{\rm kSZ}^{\ell=1}$ (60 k-modes), and ${\rm Cov}$ is the covariance matrix. We estimate $F_{\alpha\beta}$ using the measured covariance matrix as described in Appendix~\ref{app:method} and the best-fitting parameter values from Section~\ref{sec:results}. The $1\sigma$ confidence level ellipses for parameter pairs, obtained by marginalizing over the others, are derived following \cite{CoeDan2009}; these are presented in the upper-right panels of Figures~\ref{fig:corner_CMASS_all_parameters} and~\ref{fig:corner_CMASS_all_parameters_H0Omw0}.

For the cosmological observables, the joint analysis enhances the FoM by \LSH{18.9\% for the $f$--$\alpha_\parallel$ pair, 18.5\% for $f$--$\alpha_\perp$, and 8.7\% for $\alpha_\parallel$--$\alpha_\perp$. For the cosmological parameters, the improvements are 27.4\% for $H_0$--$\Omega_{\rm m}$, 25.9\% for $H_0$--$w_0$, and 20.1\% for $\Omega_{\rm m}$--$w_0$}. The greater FoM improvements for cosmological parameters compared to observables are consistent with the trends found in Section~\ref{sec:results}.

\section{The power spectrum model calculation}
\label{app:theory_calculation}

The explicit expressions for $P_{mn}$ (with $m, n = 0, 1, 2, 3, 4$) in Equation~(\ref{eq:Pgg_Pgm}) are given in \cite{Howlett2019}, which we generally follow in this work. We additionally incorporate the corrections to the $P_{02}$ and $P_{12}$ terms in \cite{Howlett2019}, as identified by \cite{Qin2025}. While readers can refer to these two papers for relevant formulas, here we rearrange perturbation terms in terms of their $\mu$ dependence, which can bring convenience when doing the theoretical multipole calculations.

We expand the density auto-power spectrum $P_{gg}$, the momentum auto-power spectrum $P_{pp}$~\citep{Howlett2019}, and the density–momentum cross-power spectrum $P_{gp}$~\citep{Qin2025b} in terms of $\mu$ as follows:
\begin{subequations}
\begin{eqnarray}
P_{gg}(k,\mu) &=& P_{gg}^{\mu^0} + P_{gg}^{\mu^2} \mu^2 + P_{gg}^{\mu^4} \mu^4 
                + P_{gg}^{\mu^6} \mu^6 + P_{gg}^{\mu^8} \mu^8\,, \\
P_{pp}(k,\mu) &=& P_{pp}^{\mu^0} + P_{pp}^{\mu^2} \mu^2 + P_{pp}^{\mu^4} \mu^4 
                + P_{pp}^{\mu^6} \mu^6\,, \\
i P_{gp}(k,\mu) &=& P_{gp}^{\mu^1} \mu + P_{gp}^{\mu^3} \mu^3 
                  + P_{gp}^{\mu^5} \mu^5 + P_{gp}^{\mu^7} \mu^7\,,
\end{eqnarray}
\end{subequations}
where
\LSH{
\begin{equation}
\begin{split} 
P_{gg}^{\mu^0}(k) &= P_{00}^{\mu^0}\,,\\
P_{gg}^{\mu^2}(k) &= 2 P_{01}^{\mu^0} + P_{02}^{\mu^0} + P_{11}^{\mu^0},\\   P_{gg}^{\mu^4}(k) &= P_{02}^{\mu^2} + P_{03}^{\mu^0} + P_{04}^{\mu^0} + P_{11}^{\mu^2} + P_{12}^{\mu^0} + P_{13}^{\mu^0} + P_{22}^{\mu^0}/4\,,\\                         
P_{gg}^{\mu^6}(k) &= P_{04}^{\mu^2} + P_{12}^{\mu^2} + P_{13}^{\mu^2} + P_{22}^{\mu^2}/4\,,\\
P_{gg}^{\mu^8}(k) &= P_{22}^{\mu^4}/4\,,\\
P_{pp}^{\mu^0}(k) &= (a H/k)^2 P_{11}^{\mu^0}\,,\\                                                   
P_{pp}^{\mu^2}(k) &= (a H/k)^2 (P_{11}^{\mu^2} + 2 P_{12}^{\mu^0} + 3 P_{13}^{\mu^0} + P_{22}^{\mu^0})\,,\\                                   
P_{pp}^{\mu^4}(k) &= (a H/k)^2 (2 P_{12}^{\mu^2} + 3 P_{13}^{\mu^2} + P_{22}^{\mu^2})\,,\\             
P_{pp}^{\mu^6}(k) &= (a H/k)^2 P_{22}^{\mu^4}\,,\\
P_{gp}^{\mu^1}(k) &= (-a H/k) (P_{01}^{\mu^0} + P_{02}^{\mu^0} + P_{11}^{\mu^0})\,,\\
P_{gp}^{\mu^3}(k) &= (-a H/k) (P_{02}^{\mu^2} + 3 P_{03}/2 + 2 P_{04}^{\mu^0} + P_{11}^{\mu^2} + 3 P_{12}^{\mu^0}/2 + 2 P_{13}^{\mu^0} + P_{22}^{\mu^0}/2)\,,\\                                  
P_{gp}^{\mu^5}(k) &= (-a H/k) (2 P_{04}^{\mu^2} + 3 P_{12}^{\mu^2}/2 + 2 P_{13}^{\mu^2} + P_{22}^{\mu^2}/2)\,,\\
P_{gp}^{\mu^7}(k) &= (-a H/k) P_{22}^{\mu^4}/2\,,
\end{split}
\end{equation}
}
with
\begin{equation}
\label{eq:P_mn_mu}
\begin{split} 
P_{00}^{\mu^0}(k) &= b_1^2 D^2 \big( P_{\text{lin}} + 2 D^2 (I_{00} + 3 k^2  P_{\text{lin}} J_{00}) \big) 
   + 2 b_1 D^4 (b_2 K_{00} + b_s K^s_{00} + b_{3,\text{nl}} \sigma_3^2  P_{\text{lin}}) \\
   &+ D^4 \big( \tfrac{1}{2} b_2^2 K_{01} + \tfrac{1}{2} b_s^2 K^s_{01} + b_2 b_s K^s_{02} \big)\,, \\
P_{01}^{\mu^0}(k) &= f b_1 D^2 \big( P_{\text{lin}} + 2 D^2 (I_{01} + b_1 I_{10} + 3 k^2  P_{\text{lin}} (J_{01} + b_1 J_{10})) 
   - b_2 D^2 K_{11} - b_s D^2 K^s_{11} \big) \\
   &- f D^4 (b_2 K_{10} + b_s K^s_{10} + b_{3,\text{nl}} \sigma_3^2  P_{\text{lin}})\,, \\
P_{02}^{\mu^0}(k) &= f^2 b_1 D^4 (I_{02} + 2 k^2  P_{\text{lin}} J_{02}) 
   - f^2 k^2 ( \sigma_{v,1}^2/f^2) P_{00}^{\mu^0} 
   + f^2 D^4 (b_2 K_{20} + b_s K^s_{20})\,, \\
P_{02}^{\mu^2}(k) &= f^2 b_1 D^4 (I_{20} + 2 k^2  P_{\text{lin}} J_{20}) 
   + f^2 D^4 (b_2 K_{30} + b_s K^s_{30})\,, \\
P_{03}^{\mu^0}(k) &= -f^2 k^2 ( \sigma_{v,2}^2/f^2) P_{01}^{\mu^0}\,, \\
P_{04}^{\mu^0}(k) &= -\tfrac{1}{2} f^4 b_1 k^2 ( \sigma_{v,1}^2/f^2) D^4 (I_{02} + 2 k^2  P_{\text{lin}} J_{02}) 
   + \tfrac{1}{4} f^4 b_1^2 k^4 P_{00}^{\mu^0} \big( ( \sigma_{v,1}^2/f^2)^2 + D^4 \sigma_4^2 \big)\,, \\
P_{04}^{\mu^2}(k) &= -\tfrac{1}{2} f^4 b_1 k^2 ( \sigma_{v,1}^2/f^2) D^4 (I_{20} + 2 k^2  P_{\text{lin}} J_{20})\,, \\
P_{11}^{\mu^0}(k) &= f^2 D^2 b_1^2 D^2 I_{31}\,, \\
P_{11}^{\mu^2}(k) &= f^2 D^2 \big( P_{\text{lin}} + D^2 (2 I_{11} + 4 b_1 I_{22} + b_1^2 I_{13} 
   + 6 k^2  P_{\text{lin}} (J_{11} + 2 b_1 J_{10})) \big)\,, \\
P_{12}^{\mu^0}(k) &= f^3 D^4 (I_{12} - b_1 I_{03} + 2 k^2  P_{\text{lin}} J_{02}) 
   - f^2 k^2 ( \sigma_{v,1}^2/f^2) P_{01}^{\mu^0} 
   + 2 f^3 k^2 D^4 ( \sigma_{v,1}^2/f^2) \big( I_{01} + I_{10}  \\
   &+ 3 k^2  P_{\text{lin}} (J_{01} + J_{10}) \big)\,, \\
P_{12}^{\mu^2}(k) &= f^3 D^4 (I_{21} - b_1 I_{30} + 2 k^2  P_{\text{lin}} J_{20})\,, \\
P_{13}^{\mu^0}(k) &= -f^4 k^2 D^2 \big( ( \sigma_{v,1}^2/f^2) b_1^2 D^2 I_{31} \big)\,, \\
P_{13}^{\mu^2}(k) &= -f^4 k^2 D^2 \big( ( \sigma_{v,2}^2/f^2) ( P_{\text{lin}} + D^2 (2 I_{11} + 4 b_1 I_{22} 
   + b_1^2 I_{13} + 6 k^2  P_{\text{lin}} (J_{11} + 2 b_1 J_{10}))) \big)\,, \\
P_{22}^{\mu^0}(k) &= \tfrac{1}{4} f^4 D^4 I_{23} 
   + f^4 k^4 ( \sigma_{v,1}^2/f^2)^2 P_{00}^{\mu^0} 
   - f^2 k^2 ( \sigma_{v,1}^2/f^2) \big( 2 P_{02}^{\mu^0} 
   - f^2 D^4 (b_2 K_{20} + b_s K^s_{20}) \big)\,, \\
P_{22}^{\mu^2}(k) &= \tfrac{1}{4} f^4 D^4 \cdot 2 I_{32} 
   - f^2 k^2 ( \sigma_{v,1}^2/f^2) \big( 2 P_{02}^{\mu^2} 
   - f^2 D^4 (b_2 K_{30} + b_s K^s_{30}) \big)\,, \\
P_{22}^{\mu^4}(k) &= \tfrac{1}{4} f^4 D^4 I_{33}\,.
\end{split}
\end{equation}
Here, $b_s$ and $b_{3,\text{nl}}$ are expressed as a function of the linear bias $b_s=-4/7(b_1-1)$ and $b_{3,\text{nl}}=32/315(b_1-1)$~\citep{Saito2014,Howlett2019}. {$D(z)$ is the the linear growth factor, $a(z)$ is the scale factor, and $P_{\text{lin}}$ is the linear matter power spectrum at $z=0$.} \LSH{In Equation~(\ref{eq:P_mn_mu}), following \cite{Howlett2019,Vlah2012}, the velocity dispersion is taken as $\sigma_{v,2}^2 / f^2$ for the components $P_{03}^{\mu^0}$, $P_{13}^{\mu^2}$, and as $\sigma_{v,1}^2 / f^2$ for the others. }

{The acquisition of $P_{mn}^{\mu^i}$ requires the direct numerical integration of $I_{nm}$, $J_{mn}$, $K^{(s)}_{mn}$, $\sigma^2_3$ and $\sigma^2_4$. Among them, the expressions for $I_{nm}$ and $J_{mn}$ can be found in Appendix D of \cite{Vlah2012}, and $K^{(s)}_{mn}$, $\sigma^2_3$, and $\sigma^2_4$ are provided in Appendix A of \cite{Howlett2019}. The direct numerical integration of them is relatively slow. Next, we introduce how to calculate them using FFTLog~\citep{Talman1978,Hamilton2000}, following the 1D Fast Fourier Transform methodology from~\cite{Schmittfull2016}.
}

First, we define the generalized 1D Hankel transform and its inverse transform:
\begin{equation}
\label{eq:kPlin}
    k^nP_{\text{lin}}(k)=4\pi\int^\infty_0 dr r^2j_\ell(kr)\xi^\ell_n\,,
\end{equation}
\begin{equation}
\label{eq:xiln}
\begin{split}
    \xi^\ell_n(r) = i^\ell \int_{\boldsymbol{q}} e^{-i\boldsymbol{q}\cdot \boldsymbol{r}} q^n \mathcal{L}_\ell(\hat{\boldsymbol{q}}\cdot\hat{\boldsymbol{r}}) P_{\text{lin}}(q) 
    = \int^\infty_0 \frac{dq}{2\pi^2} q^{2+n} j_\ell(qr) P_{\text{lin}}(q)\,.
\end{split}
\end{equation}
Here, $j_\ell$ denotes the spherical Bessel function, and $\mathcal{L}_\ell$ represents the Legendre polynomial. The quantities $k^n P_{\text{lin}}$ and $\xi^\ell_n$ form a Fourier transform pair, which can be efficiently computed using the FFTLog algorithm\footnote{\href{https://github.com/eelregit/mcfit}{https://github.com/eelregit/mcfit}}.

According to Equations.~(31) and~(40) in~\cite{Schmittfull2016}, other forms of  integral over the linear power spectrum can also be expressed by the Hankel transformation and efficiently evaluated by FFTLog, such as
\begin{equation}
\label{eq:FFTLog_1}
    \begin{split}    
    \int_{\boldsymbol{q}}q^{n_1}|\boldsymbol{k}-\boldsymbol{q}|^{n_2}\mathcal{L}_\ell(\hat{\boldsymbol{q}}\cdot (\widehat{\boldsymbol{k}-\boldsymbol{q}}))P_{\rm lin}(q)P_{\rm lin}(|\boldsymbol{k}-\boldsymbol{q}|)=(-1)^\ell4\pi\int^\infty_0drr^2j_0(kr)\xi^\ell_{n_1}(r)\xi^\ell_{n_2}(r)\,,
    \end{split}
\end{equation}
\begin{equation}
\label{eq:FFTLog_2}
    \begin{split}    
    \int_{\boldsymbol{q}}\frac{1}{|\boldsymbol{k}-\boldsymbol{q}|^{2}}q^{n}(\hat{\boldsymbol{k}}\cdot\hat{\boldsymbol{q}})^\ell P_{\rm lin}(q)=\sum^\ell_{\ell'=0}(\ell'+1)\alpha_{\ell\ell'}\int^\infty_0drr^2j_{\ell'}(kr)\xi^{\ell'}_{n}(r)/r\,,
    \end{split}
\end{equation}
with
\begin{equation}
\label{eq:alpha_ll}
\alpha_{l l'} = \frac{1}{2} \int_{-1}^{1} \mu^l \mathcal{L}_{l'}(\mu) \, d\mu = 
\begin{cases}
\displaystyle \frac{l!}{2^{(l-l')/2} \left[\frac{(l - l')}{2}\right]! (l + l' + 1)!!}, & \text{if } l \geq l' \ \& \ l \text{ and } l' \text{ both even or odd,} \\
0, & \text{otherwise.}
\end{cases}
\end{equation}

In turn, we can provide the corresponding integral equations obtainable via the Hankel transformation~\citep{Vlah2012,Howlett2019}:
\begin{equation}
\label{eq:Imn}
    \begin{split}    
    I_{mn}=\int\frac{d^3q}{(2\pi)^3}f'_{mn}(\boldsymbol{k},\boldsymbol{q})P_{\rm lin}(q)P_{\rm lin}(|\boldsymbol{k}-\boldsymbol{q}|)
    =4\pi\int^\infty_0drr^2j_0(kr)f_{mn}\,,
    \end{split}
\end{equation}
\begin{equation}
\label{eq:Kmn}
    \begin{split}    
    K^{(s)}_{mn}=\int\frac{d^3q}{(2\pi)^3}h'^{(s)}_{mn}(\boldsymbol{k},\boldsymbol{q})P_{\rm lin}(q)P_{\rm lin}(|\boldsymbol{k}-\boldsymbol{q}|)
    =4\pi\int^\infty_0drr^2j_0(kr)h^{(s)}_{mn}\,,
    \end{split}
\end{equation}
\begin{equation}
\label{eq:Jmn}
    \begin{split}    
    J_{mn}=\int\frac{d^3q}{(2\pi)^3}g'_{mn}(\frac{q}{k})\frac{P_{\rm lin}(q)}{q^2}
    =\int_{\boldsymbol{q}}\frac{1}{|\boldsymbol{k}-\boldsymbol{q}|^{2}}g_{mn} P_{\rm lin}(q)\,,
    \end{split}
\end{equation}
\begin{equation}
\label{eq:sigma32}
    \begin{split}    
    \sigma^2_3=\int_{\boldsymbol{q}}\frac{1}{|\boldsymbol{k}-\boldsymbol{q}|^{2}}s_3^2 P_{\rm lin}(q)\,,
    \end{split}
\end{equation}
\begin{equation}
\label{eq:sigma42}
    \begin{split} 
    \sigma^2_4=\frac{8}{45\pi^2}\int_0^\infty\frac{dk}{k^2}\left[4\pi\int^\infty_0drr^2j_0(kr)s^2_4\right]\,.
    \end{split}
\end{equation}
The kernels in the above integral equations can be divided into two categories -- namely $f_{mn}$, $h^{(s)}_{mn}$, $s_4^2$ as  $kernelA$ and $g_{mn}$, $s_3^2$ as $kernelB$. They can be calculated separately via
\begin{equation}
\label{eq: kernel}
    \begin{split} 
     kernelA=\sum_{\ell,n_1,n_2}A^\ell_{n_1,n_2}\xi^\ell_{n_1}\xi^\ell_{n_2}\,\,\,\,\,\,and\,\,\,\,\,\,\,\,\, kernelB=\sum_{n,\ell}B^\ell_{n}q^{n}(\hat{\boldsymbol{k}}\cdot\hat{\boldsymbol{q}})^\ell\,.
    \end{split}
\end{equation}
The coefficients of $A^\ell_{n_1,n_2}$ and $B^\ell_{n}$ are shown in table~\ref{tab: kernel1} and \ref{tab: kernel2}, respectively.

The derivations of Equations~(\ref{eq:Imn}),~(\ref{eq:Kmn}) and~(\ref{eq:sigma42}) incorporate the identity given in Equation~(\ref{eq:FFTLog_1}). The terms $I_{mn}$, $K^{(s)}_{mn}$, and $\sigma^2_4$ can be computed directly via FFTLog. For $J_{mn}$ and $\sigma_3^2$, we first express them as sums of Hankel transforms using Equation~(\ref{eq:FFTLog_2}) and then evaluate these transforms via FFTLog. \LSH{The numerical code we have developed for the model can be found in \url{https://github.com/shaohongli-code/theoretical_power_spectrum}.}

\LSH{
Furthermore, there exists a general nonperturbative relationship between the galaxy pairwise velocity power spectrum ($P_{\rm pv}$) and the galaxy density power spectrum ($P_{gg}$), shown in \cite{Sugiyama2016,Sugiyama2017}:
\begin{equation}
    \label{eq:Pv_Pgg}
    2P_{gp}=P_{\rm pv}(k,\mu)=\left(i\frac{aHf}{k\mu}\right)\frac{\partial}{\partial f}P_{gg}(k,\mu)\,.
\end{equation}
This relationship has been numerically verified in \cite{Xiao2023}. Keeping $\sigma_{v,1}^2 / f^2$ and $\sigma_{v,2}^2 / f^2$ as constants, we take the partial derivative of $P_{gg}$ in Equation~(\ref{eq:Pgg_Pgm}) with respective to $f$, according to Equation~(\ref{eq:Pv_Pgg}),  and verify the consistency between Equations~(\ref{eq:Pgg_Pgm}) and~(\ref{eq:Pv_Pgg}). The results are shown in Figure~\ref{fig:Pv}, which demonstrates that the $P_{gg}$ and $P_{gp}$ in Equation~(\ref{eq:Pgg_Pgm}) fully obey the relationship presented by Equation~(\ref{eq:Pv_Pgg}).

\begin{figure*}
\centering
\includegraphics[width=0.45\textwidth]{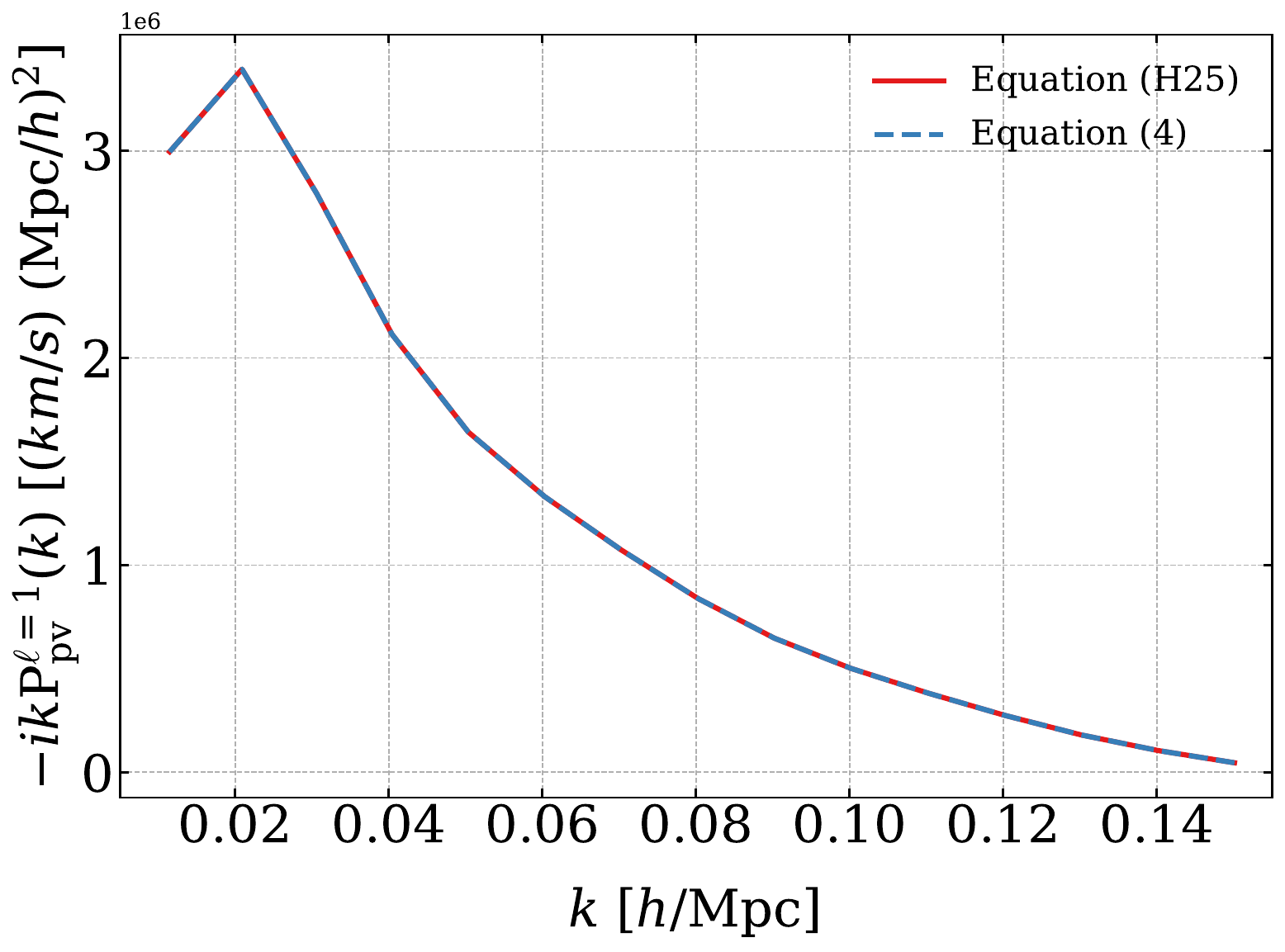}
\caption{\label{fig:Pv} Comparison between between Equation~(\ref{eq:Pgm2Ppv}) (blue dashed line) and Equation~(\ref{eq:Pv_Pgg}) (red solid line). The window function effects are included.}
\end{figure*}

}

\begin{table}
\caption{\label{tab: kernel1} 
The coefficients of $ kernelA$, $A^\ell_{n_1,n_2}$.}
\begin{ruledtabular}
\begin{tabular}{cccccccc}
&$\xi^{0}_{0}\xi^{0}_{0}$ &$\xi^{0}_{2}\xi^{0}_{-2}$ &$\xi^{1}_{1}\xi^{1}_{-1}$ &$\xi^{2}_{0}\xi^{2}_{0}$ &$\xi^{2}_{2}\xi^{2}_{-2}$ &$\xi^{3}_{1}\xi^{3}_{-1}$ &$\xi^{4}_{0}\xi^{4}_{0}$\\
\hline
$f_{00}$ & 1219/1470 & 1/6 & 62/35 & 671/1029 & 1/3 & 8/35 & 32/1715 \\
$f_{10}$ & 41/42 & 1/6 & 66/35 & 11/21 & 1/3 & 4/35 & 0 \\
$f_{02}$ & -18/35 & 0 & -2/5 & 22/49 & 0 & 2/5 & 16/245 \\
$f_{12}$ & -38/105 & 0 & -2/5 & 34/147 & 0 & 2/5 & 32/245 \\
$f_{22}$ & 11/14 & 1/6 & 62/35 & 5/7 & 1/3 & 8/35 & 0 \\
$f_{30}$ & -14/3 & 0 & -38/5 & -4/3 & -2 & -2/5 & 0 \\
$f_{13}$ & 8/3 & 0 & 4 & 1/3 & 1 & 0 & 0 \\
$f_{32}$ & -112/15 & 8/3 & -16/5 & 152/21 & -8/3 & 16/5 & 8/35 \\
$f_{01}$ & 1003/1470 & 1/6 & 58/35 & 803/1029 & 1/3 & 12/35 & 64/1715 \\
$f_{11}$ & 851/1470 & 1/6 & 54/35 & 871/1029 & 1/3 & 16/35 & 128/1715 \\
$f_{20}$ & 356/105 & 2/3 & 50/7 & 374/147 & 4/3 & 6/7 & 16/245 \\
$f_{21}$ & 292/105 & 2/3 & 234/35 & 454/147 & 4/3 & 46/35 & 32/245 \\
$f_{03}$ & 4/3 & -2/3 & 2/5 & -4/3 & 2/3 & -2/5 & 0 \\
$f_{31}$ & -1/3 & 1/3 & 0 & 1/3 & -1/3 & 0 & 0 \\
$f_{23}$ & 8/5 & 0 & 0 & -16/7 & 0 & 0 & 24/35 \\
$f_{33}$ & 168/5 & 0 & 288/5 & 96/7 & 16 & 32/5 & 24/35 \\
$h_{00}$ & 17/21 & 0 & 1 & 4/21 & 0 & 0 & 0 \\
$h_{01}$ & 1 & 0 & 0 & 0 & 0 & 0 & 0 \\
$h_{10}$ & 13/21 & 0 & 1 & 8/21 & 0 & 0 & 0 \\
$h_{11}$ & 1 & 0 & 1 & 0 & 0 & 0 & 0 \\
$h_{20}$ & -1/3 & 0 & 0 & 1/3 & 0 & 0 & 0 \\
$h_{30}$ & 5/3 & 0 & 2 & 1/3 & 0 & 0 & 0 \\
$h^s_{00}$ & 8/315 & 0 & 4/15 & 254/441 & 0 & 2/5 & 16/245 \\
$h^s_{01}$ & 4/45 & 0 & 0 & 8/63 & 0 & 0 & 8/35 \\
$h^s_{02}$ & 0 & 0 & 0 & 2/3 & 0 & 0 & 0 \\
$h^s_{10}$ & 16/315 & 0 & 4/15 & 214/441 & 0 & 2/5 & 32/245 \\
$h^s_{11}$ & 0 & 0 & 4/15 & 2/3 & 0 & 2/5 & 0 \\
$h^s_{20}$ & 2/45 & 0 & 0 & -10/63 & 0 & 0 & 4/35 \\
$h^s_{30}$ & 2/45 & 0 & 8/15 & 74/63 & 0 & 4/5 & 4/35 \\
$s^2_4$ & 47/60 & 5/12 & 11/5 & 26/21 & 1/3 & 4/5 & 8/35 \\
\end{tabular}
\end{ruledtabular}
\end{table}

\begin{table}
\caption{\label{tab: kernel2} 
The coefficients of $ kernelB$, $B^\ell_{n}$. $(n,\ell)$ is for $B_n^\ell$ or $q^{n}(\hat{\boldsymbol{k}}\cdot\hat{\boldsymbol{q}})^{\ell}.$}
\begin{ruledtabular}
\begin{tabular}{cccccccccccc}
$(n,\ell)\,{\rm of}\,B^\ell_{n}$ & $(-2,0)$ & $(0,0)$ & $(2,0)$ & $(-1,1)$ & $(1,1)$ & $(-2,2)$ & $(0,2)$ & $(2,2)$ & $(-1,3)$ & $(1,3)$ & $(0,4)$ \\
\hline
$g_{00}$ & 0 & 5/63 & 0 & -23/378 & -11/54 & -1/6 & -23/378 & 0 & 19/63 & 1/9 & 0 \\
$g_{01}$ & 0 & 1/63 & 0 & 17/378 & -37/378 & -1/6 & -55/378 & 0 & 17/63 & 5/63 & 0 \\
$g_{10}$ & -1/18 & 2/63 & -1/18 & -1/18 & -25/126 & 0 & 5/14 & 1/6 & 0 & -4/21 & 0 \\
$g_{11}$ & 0 & -1/21 & 0 & 19/126 & 1/126 & -1/6 & -29/126 & 0 & 5/21 & 1/21 & 0 \\
$g_{02}$ & 0 & -3/7 & 0 & 1/2 & 1/2 & 0 & -1/7 & 0 & -1/2 & -1/2 & 4/7 \\
$g_{20}$ & 0 & 3/7 & 0 & 5/14 & -1/2 & -1 & -12/7 & 0 & 37/14 & 3/2 & -12/7 \\
$\sigma_3^2$ & 0 & 35/24 & -5/12 & 0 & 5/6 & 0 & -5/2 & 5/4 & 0 & -5/2 & 15/8 \\
\end{tabular}
\end{ruledtabular}
\end{table}

\bibliography{mybib}{}
\bibliographystyle{aasjournal}
\end{document}